\documentclass[a4paper, USenglish]{lipics-v2021}
\pdfoutput=1

\usepackage{etoolbox}
\newtoggle{tr}
\toggletrue{tr} %

\usepackage{xspace}

\usepackage[utf8]{inputenc}
\usepackage{flushend}

\usepackage{amsmath,amssymb,amsfonts,bm,mathtools}

\usepackage{csquotes}
\usepackage{booktabs}
\usepackage{multirow}

\usepackage{pifont}

\usepackage{todonotes}
\usepackage{nicefrac}
\usepackage[noend]{algpseudocode}
\usepackage{algorithm}
\algnewcommand{\Label}[1]{\Statex \hspace{-\algorithmicindent}\hspace{-\labelsep} \textbf{#1}}
\algnewcommand{\Input}{\item[\rlap{\textbf{Input:}}{\hphantom{\textbf{Output:}}}]}
\algnewcommand{\Output}{\item[\textbf{Output:}]}
\algnewcommand{\Seq}{\item[\phantom{\textbf{Output:}}]}
\algnewcommand{\IIf}[1]{\State\algorithmicif\ #1\ \algorithmicthen}
\algnewcommand{\EndIIf}{\unskip\ \algorithmicend\ \algorithmicif}
\algnewcommand{\Downto}{\textbf{ downto }}
\algnewcommand{\By}{\textbf{ by }}
\newcommand{\cupEq}{\,\,\cup\!\!=}

\DeclareMathOperator*{\argmin}{arg\,min}

\algnewcommand\algorithmicparfor{\textbf{parfor}}
\algblockdefx{ParFor}{EndParFor}[1]
{\algorithmicparfor\ #1\ \algorithmicdo}
{\algorithmicend\ \algorithmicparfor}

\makeatletter
\ifthenelse{\equal{\ALG@noend}{t}}%
{\algtext*{EndParFor}}
{}%
\makeatother

\usepackage{siunitx}
\makeatletter
\@ifpackagelater{siunitx}{2021/05/17}{%
}{%
	\newcommand{\qty}[2]{\SI{#1}{#2}}
}
\makeatother

\iftoggle{tr}{
    \usepackage[section,above,below]{placeins}
}{
    \usepackage{placeins}
}

\usepackage{hyperref}

\newcommand {\etal}{\mbox{et al.}\xspace} %
\newcommand {\ie}{\mbox{i.\,e.}\xspace}     %
\newcommand {\eg}{\mbox{e.\,g.}\xspace}     %

\newcommand{\tabref}[1]{Table\;\ref{#1}\xspace}
\newcommand{\figref}[1]{Figure\;\ref{#1}\xspace}
\newcommand{\secref}[1]{Section\;\ref{#1}\xspace}

\renewcommand{\algref}[1]{Algorithm\;\ref{#1}\xspace}

\newcommand{\alglnref}[2]{Algorithm\;\hyperref[#1:#2]{\ref*{#1}::\ref*{#1:#2}\xspace}}

\makeatletter
\newcommand{\eqnref}{\@ifstar
    \eqnrefS%
    \eqnrefNS%
}
\makeatother
\newcommand{\eqnrefNS}[1]{Equation\;\eqref{#1}\xspace}
\newcommand{\eqnrefS}[1]{Equation\;\ref{#1}\xspace}

\newcommand{\set}[1]{\ensuremath{\mathbf{#1}}}
\newcommand{\prj}[2]{\ensuremath{\rho_{#2}\left(#1\right)}}

\newcommand{\Oh}[1]{\ensuremath{\mathcal{O}\!\left(#1\right)}\xspace}

\newcommand{\Oone}{\Oh{1}}
\newcommand{\Ologn}{\Oh{\log n}}
\newcommand{\On}{\Oh{n}}
\newcommand{\Onlogn}{\Oh{n \log n}}
\newcommand{\OnSq}{\Oh{n^2}}
\newcommand{\OnSqrt}{\Oh{\sqrt{n}}}

\newcommand{\GitUrl}{\url{https://github.com/dfunke/YaoGraph}}

\title{Efficient Yao Graph Construction} %

\author{Daniel Funke}{Karlsruhe Institut of Technology}{funke@kit.edu}{}{}%

\author{Peter Sanders}{Karlsruhe Institut of Technology}{sanders@kit.edu}{}{}

\authorrunning{D. Funke and P. Sanders} %

\Copyright{Daniel Funke and Peter Sanders} %



\category{} %

\supplement{Source code available at \GitUrl}%

\acknowledgements{}%

\nolinenumbers %

\iftoggle{tr}{
\hideLIPIcs %
}

\begin{document}
    
\maketitle

\begin{abstract}
Yao graphs are geometric spanners that connect each point of a given point set to its nearest neighbor in each of $k$ cones drawn around it.
Yao graphs were introduced to construct minimum spanning trees in $d$ dimensional spaces.
Moreover, they are used for instance in topology control in wireless networks.
An optimal \Onlogn time algorithm to construct Yao graphs for given point set has been proposed in the literature
but -- to the best of our knowledge -- never been implemented.
Instead, algorithms with a  quadratic complexity are used in popular packages to construct these graphs.
In this paper we present the first implementation of the optimal Yao graph algorithm. 
We develop and tune the data structures required to achieve the \Onlogn bound
and detail algorithmic adaptions necessary to take the original algorithm from theory to practice.
We propose a priority queue data structure that separates static and dynamic events 
and might be of independent interest for other sweepline algorithms.
Additionally, we propose a new Yao graph algorithm based on a uniform grid data structure that performs well for medium-sized inputs.
We evaluate our implementations on a wide variety synthetic and real-world datasets and show
that our implementation outperforms current publicly available implementations by at least an order of magnitude.
\end{abstract}

\section{Introduction}\label{sec:intro}

\begin{figure}[tb]
    \centering
    \includegraphics[width=.33\textwidth]{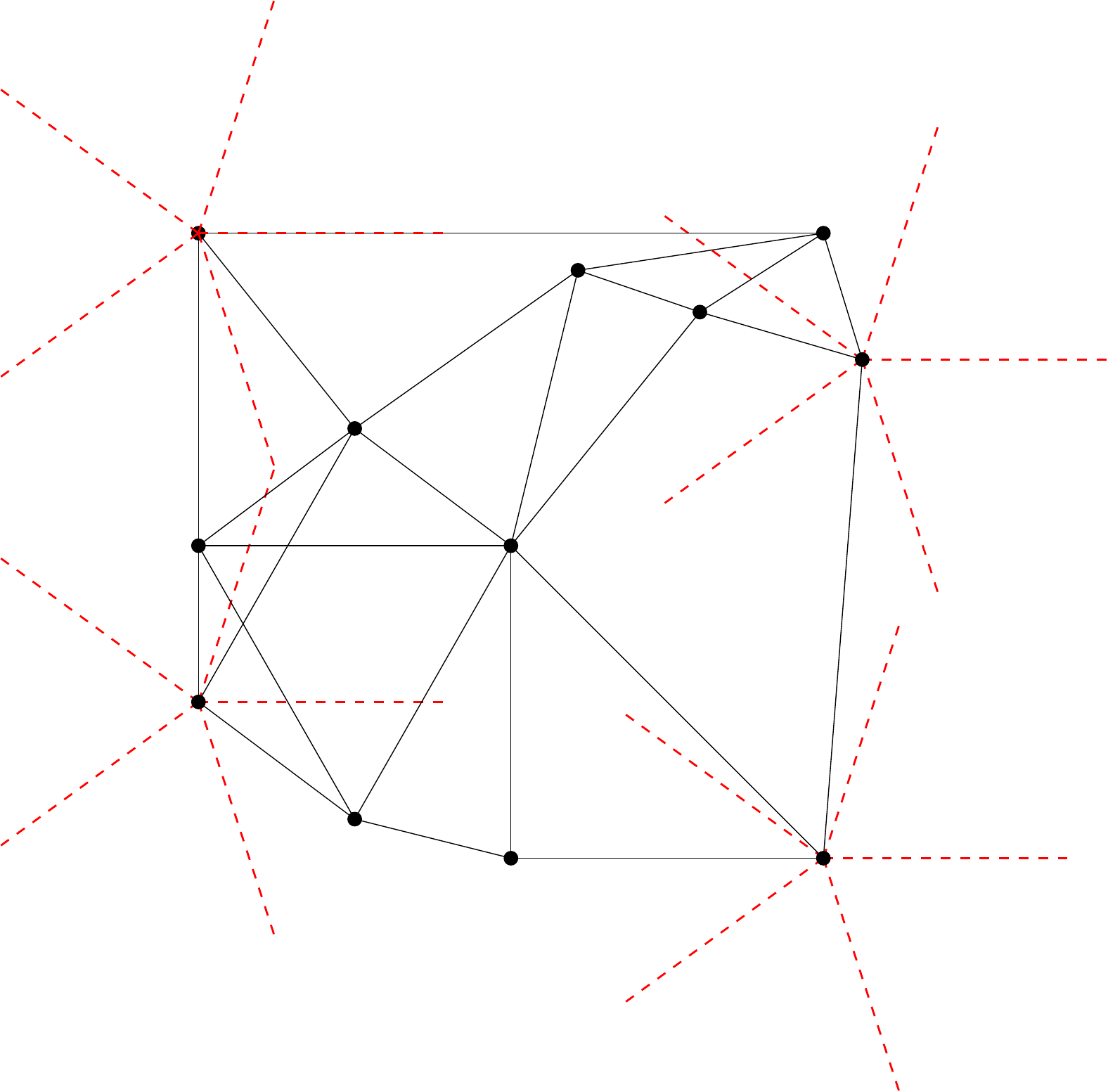}
    \caption{Yao graph for ten points and $k = 5$ cones.
        The five cones are illustrated as red dashed lines around four example points.}
    \label{fig:yao_graph}
\end{figure}

Yao graphs are geometric spanners that connect each point of a given point set to its nearest neighbor in each of $k$ cones, refer to \figref{fig:yao_graph} for an example.
A $t$-spanner is a weighted graph, where for any pair of vertices there exists a $t$-path between them with weight at most $t$ times their spatial distance.
Parameter $t$ is known as the \emph{stretch factor} of the spanner.
For $k > 6$, the stretch factor of  Yao graphs can be bounded by $\frac{1}{1 - 2 \sin{\nicefrac{\pi}{k}}}$ \cite{ThetaSL}.
Yao introduced these kind of graphs to construct minimum spanning trees in $d$ dimensional space \cite{YaoMST}.
Moreover, they are used for instance in topology control in wireless networks \cite{YaoWireless}. 
Chang \etal \cite{YaoOpt} present an optimal algorithm to construct these graphs in \Onlogn time.
Due to the intricate nature of their algorithm 
and the reliance on expensive geometric constructions,
there is no -- to the best of our knowledge -- implementation of their algorithm available.
Instead, an algorithm with an inferior \OnSq bound is used in the cone-based spanners package of the popular
CGAL library \cite{CGAL:Cone}.

\subparagraph*{Contribution.}
In this paper we present the first publicly available implementation of Chang \etal's optimal algorithm for Yao graph construction.
We take their algorithm from theory to practice by 
developing the data structures required to achieve the \Onlogn bound
and provide detailed descriptions of all operations of the algorithm that are missing in the original paper,
such as input point ordering, handling of composite boundaries and enclosing region search.
In our event queue, we separate static (input point) events and dynamic (intersection point) events.
This greatly improves the efficiency of priority queue operations and might be a useful technique for other sweepline algorihtms.
We test our algorithm on a wide range of synthetic and real-world datasets.
We show that despite the intricate nature of the algorithm and the use of expensive geometric constructions 
our implementation achieves a speedup of an order of magnitude over other currently available implementations.
Additionally, we develop a new Yao graph algorithm based on a uniform grid data structure,
that is simple to implement, easy to parallelize and performs well for medium-sized inputs.

\subparagraph*{Outline.}
In \secref{sec:rw} we review related work on the construction of Yao graphs.
\secref{sec:algo} presents the optimal algorithm of Chang \etal 
and algorithmic adaptions necessary for its implementation.
Further implementation details such as data structures and geometric operations are described in \secref{sec:impl}.
We evaluate our implementation and compare against its competitors in \secref{sec:eval}.
\secref{sec:outro} summarizes our paper and presents an outlook on future work.

\subparagraph*{Definitions.}\label{sec:def}
Given a set $\set{P}$ of points in two-dimensional Euclidian space and an integer parameter $k > 1$,
the Yao graph $G_k = (\set{P}, \set{E})$ is a directed graph,
connecting every point $p \in \set{P}$ with its nearest neighbor in each of $k$ cones.
Every cone $\mathcal{C}_i = (\theta_L, \theta_R)$, $0 \leq i < k$, is defined by its two limiting rays 
with angles $\theta_L =  \frac{2 i \pi}{k}$ and $\theta_R = \frac{2 (i+1) \pi}{k}$. 
We denote $\mathcal{C}_i^p$ as the cone $\mathcal{C}_i$ with apex at point $p$.
We furthermore define, that the \emph{left} -- or counterclockwise -- boundary ray with angle $\theta_L$ belongs to a cone $\mathcal{C}$,
whereas the \emph{right} one does not,
\ie for a given point $p \in \set{P}$ and cone $\mathcal{C}_i^p$ we define the set of points 
$\set{P} \cap \mathcal{C}_i^p := \{q \in \set{P}:  \sphericalangle p \, q \in [\theta_L, \theta_R) \}$.
Then the edge set $\set{E}$ of  the Yao graph $G_k = (\set{P}, \set{E})$ can be formally defined as 
$\set{E} := \{\forall i \in [0,k), \forall p \in \set{P}: (p,q) \text{ with } q = \argmin_{v \in \set{P} \cap \mathcal{C}_i^p} \left( d(p,v) \right)  \}$,
with $d(\cdot, \cdot)$ denoting the Euclidean distance function.

\section{Related Work}\label{sec:rw}
Yao presents an $\Oh{n^{\nicefrac{5}{3}} \log n}$ time algorithm to compute a solution to the 
\emph{Eight-Neighbors Problem} -- a Yao graph with $k = 8$.
It is based on a tessellation of the Euclidean space into cells. 
For a given point and cone, each cell is characterized whether it can contain nearest-neighbor candidates
to reduce the number of necessary distance computations.
The problem is solved optimally by Chang \etal,
who present a \Onlogn time algorithm for
constructing the Yao graph of a given point set and parameter $k$ \cite{YaoOpt}.
Their algorithm follows the same structure as Fortune's algorithm for constructing the
Voronoi diagram of a point set \cite{FortuneSL},
using the sweepline technique originally introduced by  Bentley and Ottmann for computing line-segement intersections \cite{BenOtt79}.
However, even though there are many implementations of Fortune's algorithm available,
there is no implementation of Chang \etal's Yao graph algorithm that we are aware of.
Instead, for instance the CGAL library's cone-based spanners package implements a less efficient \OnSq time algorithm \cite{CGAL:Cone}.
Their algorithm is an adaption of a sweepline algorithm for constructing $\Theta$-graphs \cite{ThetaSL}.
$\Theta$-graphs are defined similarly to Yao graphs,
except that the nearest neighbor in each cone is not defined by Euclidean distance
but by the projection distance onto the cone's internal angle bisector.
This allows for a \Onlogn sweepline algorithm,
that uses a balanced search tree as sweepline data structure to answer one-dimensional range queries \cite{ThetaSL}.
For Yao graphs, such a reduction in dimensionality is not possible,
thus, CGAL's algorithm employs linear search within the sweepline data structure to find the nearest neighbor, 
leading to the \OnSq bound.
However, CGAL's algorithm is much simpler to implement than the optimal algorithm proposed by \cite{YaoOpt}
and does not require geometric constructions, just predicates.
\tabref{tab:algComp} in \secref{sec:impl} provides an overview of the required geometric operations by both algorithms.
 
Applications of Yao graphs in wireless networks require different techniques and algorithms,
as each point -- or node  -- within the network has only a partial view of all available sites.
Therefore, Yao graphs of these networks need to be computed locally by the nodes with their partial information.
Zhang \etal present an algorithm for this setting, which also takes disturbances from radio interference into account \cite{YaoLocalWireless}.

\section{Algorithm}\label{sec:algo}

\begin{figure}[tb]
    \centering
    \includegraphics[width=.7\textwidth]{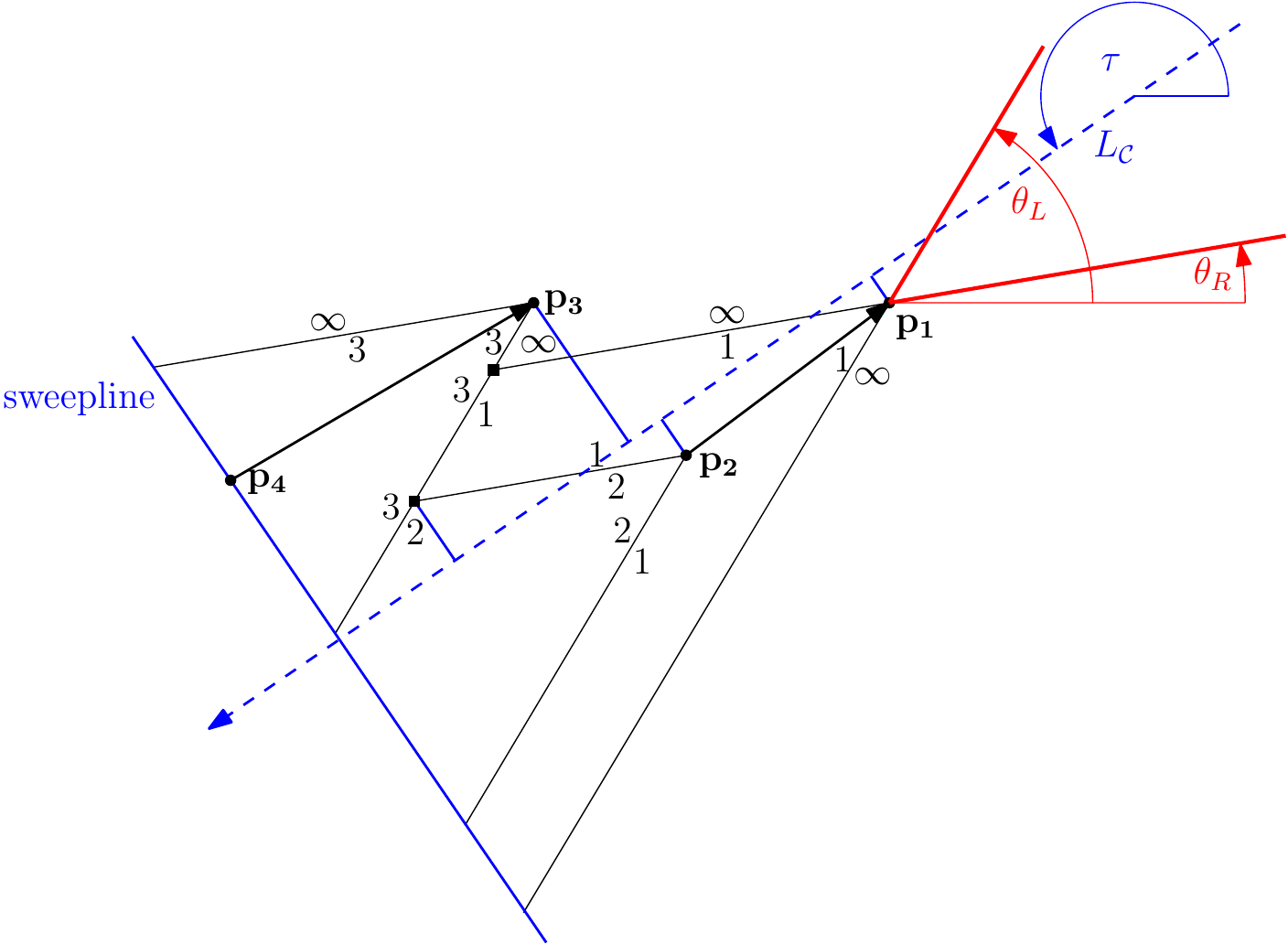}
    \caption{Example state of the sweepline algorithm for cone $\mathcal{C} = (\theta_L, \theta_R)$ (marked in red). 
        Input points (circles, bold labels) are numbered in the order they are swept by the sweepline,
        with their projections on the cone's internal angle bisector shown in blue.
        Rays are labeled with the regions they are separating.
        Intersection events are marked with a square.
        Already determined edges of the Yao graph are indicated by arrows.}
    \label{fig:sweepline_state}
\end{figure}

\begin{algorithm}[tbp]
	\begin{algorithmic}[1]
		\Input points $\set{P} = \{p_1, \dots, p_n \}$ with $p_i \in \mathbb{R}^2$, cone $(\theta_L, \theta_R)$
		\Output GNG $G = (V, E)$
		
		\medskip
		
		\State $\tau \gets \frac{\theta_L + \theta_R}{2} + \pi$ \Comment opposite of cone's internal angle bisector
		\State $Q \gets \left\{ (\prj{p}{\tau} ,p, I): p \in \set{P} \right\}$ \Comment initialize PQ with input points
		\State $SL \gets \emptyset$
		\State $G = (V, E) \gets (\set{P}, \emptyset)$
		
		\medskip
		
		\While{$p \gets \operatorname{popMin}(Q)$}
		\If{$p$ is input point}
		\State $B_L, B_R \gets \operatorname{findRegion}(p, SL)$ \Comment $B_L$ and $B_R$ enclose $p$ \label{alg:gng:find}
		
		\State $E \cupEq (p, B_L^r)$ \Comment $\operatorname{assert}\left(B_L^r == B_R^l\right)$
		
		\If{$B_L \cap B_R = v$} delete $v$ from $Q$ \label{alg:gng:checkIS}
		\EndIf
		\State $B^\ast_L \gets L\left(p, \theta_L + \pi, B_L^r, R_p\right)$
		\State $B^\ast_R \gets L\left(p, \theta_R + \pi, R_p, B_R^l\right)$
		\State insert $[B^\ast_L, B^\ast_R]$ into $SL$ between $B_L$ and $B_R$ \label{alg:gng:insert}
		\If{$B_L \cap B^\ast_L = v$} $Q \cupEq (\prj{v}{\tau}, v)$ \label{alg:gng:createIS}
		\EndIf
		\If{$B_R \cap B^\ast_R = v$} $Q \cupEq (\prj{v}{\tau}, v)$
		\EndIf
		\EndIf
		\If{$p$ is intersection}
		\State $B_L, B_R \gets \text{intersecting rays at $p$}$
		\State $a \gets B_L^l \qquad b \gets B_R^r$
		\If{$B_L \cap \operatorname{prev}(B_L) = v$} delete $v$ from $Q$ \Comment left neighbor boundary on $SL$
		\EndIf
		\If{$B_R \cap \operatorname{succ}(B_R) = v$} delete $v$ from $Q$ \Comment right neighbor boundary on $SL$
		\EndIf
		\State $L_{BS} \gets L(\frac{\vec{a} + \vec{b}}{2}, \sphericalangle a b + \frac{\pi}{2}, R_a, R_b)$ 
		\Comment bisector of line segment $\overleftrightarrow{a b}$
		\State $L_L \gets L(p, \theta_L + \pi, R_a, R_b)$
		\State $L_R \gets L(p, \theta_R + \pi, R_a, R_b)$
		\If{$L_L \cap L_{BS} = \emptyset = L_{BS} \cap L_R$} \Comment bisector intersects no line from $p$
		\State $B^\ast = L\left(p, \pi + \begin{cases} \theta_L  & \text{if $\prj{a}{\tau} < \prj{b}{\tau}$} \\ \theta_R & \text{else} \end{cases}, R_a, R_b\right)$
		\EndIf
		\If{$L_L \cap L_{BS} = p = L_{BS} \cap L_R$} \Comment bisector intersects both lines in $p$
		\State $B^\ast = L(p, \sphericalangle a b + \frac{\pi}{2}, R_a, R_b)$
		\EndIf
		\If{$L_L \cap L_{BS} = v$ or  $L_{BS} \cap L_R = v$} \Comment bisector intersects one line in $v$
		\State $B^\ast = L(\overleftrightarrow{p v}, R_a, R_b) + L(v, \sphericalangle a b + \frac{\pi}{2}, R_a, R_b)$ \label{alg:gng:boundaryExt}
		\State $Q \cupEq (\prj{v}{\tau}, v)$ \Comment deletion event
		\EndIf
		\State replace $[B_L, B_R]$ in $SL$ with $B^\ast$ \label{alg:gng:replace}
		\If{$B^\ast \cap \operatorname{prev}(B^\ast) = v$} $Q \cupEq (\prj{v}{\tau}, v)$
		\EndIf
		\If{$B^\ast \cap \operatorname{succ}(B^\ast) = v$} $Q \cupEq (\prj{v}{\tau}, v)$
		\EndIf
		\EndIf
		\If{$p$ is deletion point}
		\State $B \gets \text{ray belonging to $p$}$ \Comment $B = L(\overleftrightarrow{p v}, R_a, R_b) + L(v, \sphericalangle a b + \frac{\pi}{2}, R_a, R_b)$
		\State replace $B$ in $SL$ with $L(v, \sphericalangle a b + \frac{\pi}{2}, R_a, R_b)$
		\EndIf
		\EndWhile
		
		\State \Return $G$
	\end{algorithmic}
	\caption{Sweepline algorithm for cone defined by $(\theta_L, \theta_R)$.
	$L(p, \theta, R_a, R_b)$ and $L(\overleftrightarrow{p v}, R_a, R_b)$ denote the ray originating at $p$ with angle $\theta$ and the line segment from $p$ to $v$, respectively, 
	both separating regions $R_a$ and $R_b$.}
	\label{alg:gng}
\end{algorithm}

In their 1990 paper, Chang \etal present an \Onlogn time sweepline algorithm to compute the \emph{oriented Voronoi diagram} of a point set.
Through a small modification, their algorithm can compute the \emph{geographic neighborhood graph} -- or Yao graph -- 
of a point set within the same, optimal, bound \cite[Theorem 3.2, Theorem 4.1]{YaoOpt}.\footnote{%
Our implementation computes the Yao graph but can easily be modified to compute the oriented Voronoi diagram.}
To construct the Yao graph $G_k = (\set{P}, \set{E})$ with $k$ cones for point set $\set{P}$,
$k$ sweepline passes are required,
each considering a specific cone $\mathcal{C} = (\theta_L, \theta_R)$.
The sweepline for a cone $\mathcal{C}$ proceeds in the opposing direction to the cone's internal angle bisector,
represented as line $L_\mathcal{C}$ through the origin with angle $\tau = \frac{\theta_L + \theta_R}{2} + \pi$.
Refer to the blue dashed line in \figref{fig:sweepline_state}.
Input points are swept in the order of their projection onto $L_\mathcal{C}$,
given by sorting
\begin{equation} \label{eq:proj}
	\prj{p}{\tau} := \frac{\vec{p} \cdot \vec{SL}}{\vec{SL} \cdot \vec{SL}} \quad \forall p \in \set{P} \text{ with } \vec{p} = (x,y) \text{ and } \vec{SL} = \left(\cos{\tau}, \sin{\tau}\right).
\end{equation}
All input points are inserted into an event priority queue $Q$ with priority
$\prj{p}{\tau}$ and event type \emph{input point}.
Each input point $p$ is the origin of a cone $\mathcal{C}^p$ with boundary rays $B_L$ and $B_R$.
Cone $\mathcal{C}^p$ defines the \emph{region} $R_p$ of the plane,
where point $p$ is the nearest-neighbor with respect to cone $\mathcal{C}^p$ for any point being swept after $p$.
The invariant of the algorithm is that once a point has been swept, 
its nearest neighbor in cone $\mathcal{C}$ has been determined.
For instance, in \figref{fig:sweepline_state} $p_2$ is in the region of $p_1$,
therefore $p_1$ is the nearest neighbor for $p_2$ with respect to cone $\mathcal{C}$.
A boundary ray $B_\Box$ always separates the regions of two input points,
thus it is defined by its point of origin $B_\Box^p$ and angle $B_\Box^\theta$ as well as
its left and right  region, $B_\Box^l$ and $B_\Box^r$ respectively.
The region outside any point`s cone is labeled with infinity.
The algorithm maintains an ordered data structure $SL$ of rays currently
intersecting the sweepline.
The rays are ordered left-to-right 
and the data structure needs to support insert, remove and find operations
as well as access to the left and right neighbors of a given ray.
The details of the data structure are presented in \secref{sec:sl_ds}.
An example execution of our algorithm is depicted in \figref{fig:app:example} in the appendix.

\subsection{Event Types}\label{sec:events}

There are three \emph{event types}: 1) input points, 2) intersection points, and 3) deletion points\footnote{%
Deletion points are not present in the original algorithm description by Chang \etal \cite{YaoOpt}.}.
In the following, we give an overview how each event is handled by the algorithm.
The details can be found in \algref{alg:gng}.
If several events coincide, their processing order can be arbitrary.

\subparagraph*{1) Input points.}
All points of the given set $\set{P}$ are inserted into the event priority queue at the beginning of the algorithm.
For an input point event with associated point $p$,
the sweepline data structure $SL$ is searched for the region $R_q$ containing $p$.
This region is defined by its two bounding rays $B_L$ and $B_R$ and their
associated regions $B_L^r = B_R^l = R_q$.\footnote{%
Note that Chang \etal \cite{YaoOpt} explicitly store rays and regions in their sweepline data structure.
However, since a region is definitively identifiable by its two bounding rays,
we choose this simpler representation of the sweepline state.}
We can then add edge $(p, q)$ to the edge set $\set{E}$ of $G_k$,
as proven in \cite[Lemma 3.1]{YaoOpt}.
The point $p$ is the apex of region $R_p$,
bounded to the left by $B_L^\ast = (p, \theta_L + \pi)$,
separating regions $R_q$ and $R_p$,
and bounded to the right by $B_R^\ast = (p, \theta_R + \pi)$,
separating regions $R_p$ and $R_q$.
These new rays are inserted into $SL$ between $B_L$ and $B_R$,
forming the sequence $[B_L, B_L^\ast, B_R^\ast, B_R]$.
Lastly, intersection points of the considered rays need to be addressed.
If $B_L$ and $B_R$ intersect in point $v$,
its associated intersection point event needs to be removed from the priority queue $Q$,
as $B_L$ and $B_R$ are no longer neighboring rays.
Instead, possible intersection points between $B_L$ and $B_L^\ast$ 
as well as $B_R^\ast$ and $B_R$ are added to $Q$ for future processing.

\subparagraph*{2) Intersection points.}
An intersection point $v$ is associated with its two intersecting rays $B_L$ and $B_R$.
They separate regions $R_p := B_L^l$, $B_L^r = R_m = B_R^l$ and  $B_R^r =: R_q$,
refer to \figref{fig:boundary}. 
Region $R_m$ terminates at intersection point $v$ and 
a new boundary $B^\ast$ between $R_p$ and $R_q$ originates at $v$.
The shape of $B^\ast$ depends on the configuration of points $p$ and $q$
and can either be a simple ray or a union of a line segment and a ray.
\secref{sec:boundary} describes in more detail how $B^\ast$ is determined.
$B^\ast$ then replaces the sequence $[B_L, B_R]$ in the sweepline data structure.
Again, intersection points of the considered rays need to be addressed.
If $B_L$ has an intersection point with its left neighbor or
if $B_R$ has an intersection point with its right neighbor,
the associated intersection point events need to be removed from the event queue $Q$.
Correspondingly, if $B^\ast$ intersects its left or right neighbor,
the appropriate intersection point events are added to $Q$.

\subparagraph*{3) Deletion event.}
Deletion events are not part of the original algorithm described by Chang \etal \cite{YaoOpt},
as the authors do not specify how to handle composite boundaries in their paper.
We use them to implement boundaries consisting of a line segment and a ray.
The deletion event marks the end of the line segment and the beginning of the ray.
It does not change the actual state of the sweepline data structure.

\subsection{Boundary Determination}\label{sec:boundary}

\begin{figure}[tb]
	\centering
	\begin{subfigure}[t]{.44\textwidth}
		\centering
		\includegraphics[page=1,width=\textwidth,trim=1pt 1pt 1pt 1pt,clip]{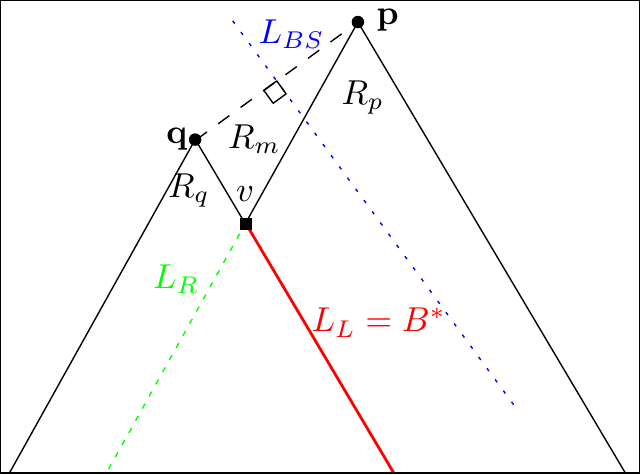}
		\caption{$L_{BS}$ intersects neither $L_L$ nor $L_R$.}
	\end{subfigure}\hfill
	\begin{subfigure}[t]{0.44\textwidth}
		\centering
		\includegraphics[page=2,width=\textwidth,trim=1pt 1pt 1pt 1pt,clip]{img/boundary}
		\caption{$L_{BS}$ intersects $L_R$ in $w$.}
	\end{subfigure}\hfill
	\begin{subfigure}[t]{0.44\textwidth}
		\centering
		\includegraphics[page=3,width=\textwidth,trim=1pt 1pt 1pt 1pt,clip]{img/boundary}
		\caption{$L_{BS}$ intersects both $L_L$ and $L_R$ in $v$.}
	\end{subfigure}\hfill
	\begin{subfigure}[t]{0.44\textwidth}
		\centering
		\includegraphics[page=4,width=\textwidth,trim=1pt 1pt 1pt 1pt,clip]{img/boundary}
		\caption{Input points lying on cone boundaries.}
		\label{fig:ColinearBoundary}
	\end{subfigure}
	\caption{Illustration of the three possible configurations for boundary $B^\ast$ following an intersection point event.
		In all examples input point $p$ is swept before $q$.
		Lines $L_L$, $L_R$ and $L_{BS}$ are dotted,
		the resulting boundary $B^\ast$ is denoted in bold.
		In Figure d), Yao graph edges are shown as bold arrows.}
	\label{fig:boundary}
\end{figure}

As described in the previous section,
at an intersection point event $v$,
the two intersecting rays $B_L$ and $B_R$ are merged into a new boundary $B^\ast$,
separating regions $R_p := B_L^r$ and $R_q := B_R^l$.
The shape of the boundary is determined by the configuration of points $p$ and $q$.
The determination is based on the number of intersection points between lines
\begin{itemize}
\item $L_L = (v, \theta_L + \pi)$ (green),
\item $L_R = (v, \theta_R + \pi)$ (red), and
\item the bisector $L_{BS}$ of line $\vec{pq}$, $L_{BS} = (\frac{\vec{p} + \vec{q}}{2}, \sphericalangle p q + \nicefrac{\pi}{2})$ (blue, dashed).
\end{itemize}
The colors refer to the lines in \figref{fig:boundary} which illustrates the different cases.
If $L_{BS}$ intersects neither $L_L$ nor $L_R$ then $B^\ast = L_L = (v, \pi + \theta_R)$ if $p$ was swept before $q$,
otherwise $B^\ast = L_R = (v, \pi + \theta_R)$.
Intuitively, the region of the lower point with respect to the sweepline direction continues,
whereas the upper region stops at intersection point $v$.
If $L_{BS}$ intersects both lines $L_L$ and $L_R$,
then the two intersection points must coincide with $v$.
In this case, $B^\ast = (p, \sphericalangle p q+ \nicefrac{\pi}{2})$,
\ie the boundary continues with the angle of the bisector from point $v$.
Otherwise, \ie $L_{BS}$ intersects either $L_L$ or $L_R$ in a point $w$,
the resulting boundary $B^\ast$ will be the union of the line segment $\vec{v w}$ and
ray $(w, \sphericalangle p q+ \nicefrac{\pi}{2})$.
In this case, a deletion point event is added to the priority queue at point $w$.
All three cases are depicted in \figref{fig:boundary}.

\subparagraph*{Input Points Colinear on Cone Boundaries.}
Chang \etal make the assumption that no line between two input points is with angle $\theta_L$ or $\theta_R$.
In the following we shall lift this requirement.
Input points sharing a common line with angle $\theta_\Box$ become \emph{visible} from each other.
This  impacts the regions the passing boundary rays are separating.
Refer to \figref{fig:ColinearBoundary} for a graphical representation of the following discussion.
Recall, that the \emph{left} -- or counterclockwise -- boundary ray with angle $\theta_L$ belongs to a cone $\mathcal{C}$,
whereas the \emph{right} one does not.
If a boundary ray $B_\Box = (p, \theta_\Box, B_\Box^l, B_\Box^r)$ intersects an input point $q$
then $B_\Box$ terminates at $q$ and a new boundary ray $B_\Box^\prime$ is formed.
If $B_\Box$ is a left boundary, \ie $\theta_\Box = \theta_L$,
then edge $(p,q)$ is added to $G_K$ and $B_\Box^\prime$ separates regions $B_\Box^l$ and $R_q$.
If it is a right boundary then no edge is added to $G_k$ and $B_\Box^\prime$ separates $R_q$ and $B_\Box^r$.
Refer to example points $q$ and $w$ in \figref{fig:ColinearBoundary}.

\subsection{Analysis}
The total number of events processed by the sweepline algorithm is the sum of input point events $N_\text{input}$,
intersection point events $N_\text{IE}$ and deletion events $N_\text{DE}$.
In order to bound the number of events processed by the sweepline algorithm,
we consider the number of rays that can be present in the sweepline data structure during the execution of the algorithm.

Every input point event adds two rays to the sweepline data structure $SL$, resulting in a total of $2 n$ rays.
It possibly removes one intersection event from the event queue $Q$ and may add up to two new such events.
Every intersection point event removes two rays from $SL$ and adds one new ray,
thus reducing the sweepline size by one.
Therefore, at most $2 n$ intersection events can be processed before all rays are removed from the sweepline.
An intersection event possibly removes two additional intersection events aside from itself from $Q$ and may add one new intersection event.\footnote{%
	Technically, $B^\ast$ can intersect both its neighbors, leading to two intersection events.
	However, when the first -- as defined by $\prj{\cdot}{\tau}$ -- intersection event is processed,
	it will delete the second event from $Q$,
	as $B^\ast$ is removed from $SL$ and a new boundary ray is inserted by the first event.}
Additionally, one deletion event may be added to $Q$.
Therefore, at most $2n$ deletion events may be processed, 
each of which leaves the number of rays and intersections unchanged.
In total,
\begin{align}\label{eq:app:nevents}
	N_\text{events} &\leq N_\text{input} + N_\text{IE} + N_\text{DE} \nonumber \\
	&\leq n + 2 n + 2 n = 5 n
\end{align}
With the appropriate data structures, each event can be processed in \Ologn time,
yielding the bound of \Onlogn. 

\section{Implementation Details}\label{sec:impl}

In this section we highlight some of the design decisions of our implementation of Chang \etal's Yao graph algorithm.

\subsection{Geometric Kernels}\label{sec:kernels}

\begin{table}[tbh]
	\centering
	\caption{Comparison of geometric predicates used in algorithms for Yao graph construction}
	\label{tab:algComp}
	\begin{tabular}{lcccc}
		\toprule
		& Chang \etal & CGAL  & Naive & Grid  \\ \cmidrule(lr){2-2}\cmidrule(lr){3-3}\cmidrule(lr){4-4}\cmidrule(lr){5-5}
		\textbf{Complexity}    &   \Onlogn   & \OnSq & \OnSq & \OnSq \\
		\textbf{Predicates}    &             &       &       &       \\
		Eucl. distance comp.   &      X      &   X   &   X   &   X   \\
		dist. to line comp.    &      X      &   X   &       &       \\
		oriented side of line  &      X      &       &  (X)  &  (X)  \\
		\textbf{Constructions} &             &       &       &       \\
		cone boundaries        &      X      &   X   &   X   &   X   \\
		box construction       &             &       &       &   X   \\
		line projection        &      X      &       &       &       \\
		ray intersection       &      X      &       &       &       \\ \bottomrule
		&             &       &       &
	\end{tabular}
\end{table}

The algorithm by Chang \etal \cite{YaoOpt} requires many different geometric predicates and constructions.
We implement our own version of the required predicates and constructions in an \emph{inexact} manner.
Additionally, the user can employ kernels provided by the CGAL library. 
The EPIC kernel provides \underline{e}xact \underline{p}redicates and \underline{i}nexact \underline{c}onstructions,
whereas the EPEC kernel features \underline{e}xact \underline{p}redicates and \underline{e}xact \underline{c}onstructions \cite{CGAL:Kernel}.

\tabref{tab:algComp} lists the geometric predicates used by the different algorithms for Yao graph construction presented in this paper.
If the boundary rays of the cones are constructed exactly,
then the naive as well as grid-based Yao graph algorithm require an oriented side of line predicate
to determine the cone $\mathcal{C}_I^p$ a point $q$ lies in with respect to point $p$.
Additionally, the grid algorithm could construct the grid data structure using exact computations.
However, in our implementation we only use inexact computations to place the input points into grid cells,
as we did not encounter the need for exact computation in any of our experiments.
Note that the determination whether a grid cell could hold a closer neighbor than the currently found one
is done using the (exact) Euclidean distance comparison predicate.

In order to reduce the number of expensive ray-ray intersection calculations,
we store all found intersection points in a hash table,
with the two intersecting rays as key,
see \eg \alglnref{alg:gng}{createIS}.
If we need to check whether two rays are intersecting -- \eg \alglnref{alg:gng}{checkIS} --
we merely require a hash table lookup.

\subsection{Sweepline Data Structure}\label{sec:sl_ds}

Chang \etal prove an \Onlogn time complexity for the algorithm~\cite[Theorem 3.2]{YaoOpt}.
In order to  achieve this bound, 
the data structure maintaining the rays currently intersected by the sweepline must provide the following operations:
insert, remove and predecessor search in \Ologn as well as left and right neighbor access in \Oone.\footnote{%
The \Onlogn bound would still hold with neighbor access in \Ologn.}

\begin{figure}[tb]
	\centering
	\begin{minipage}[t]{.44\textwidth}
		\centering
		\includegraphics[page=3, width=\textwidth]{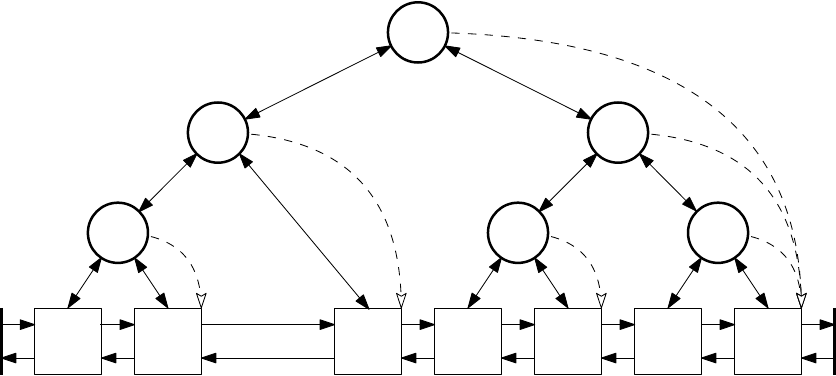}
		\caption{The sweepline data structure is a doubly linked list of rays with an AVL search tree on top.
			Additionally, each node has a pointer to the rightmost ray in its subtree (dashed)}
		\label{fig:sweepline_ds}
	\end{minipage}\hfill
	\begin{minipage}[t]{.44\textwidth}
		\centering
		\includegraphics[page=4, width=\textwidth]{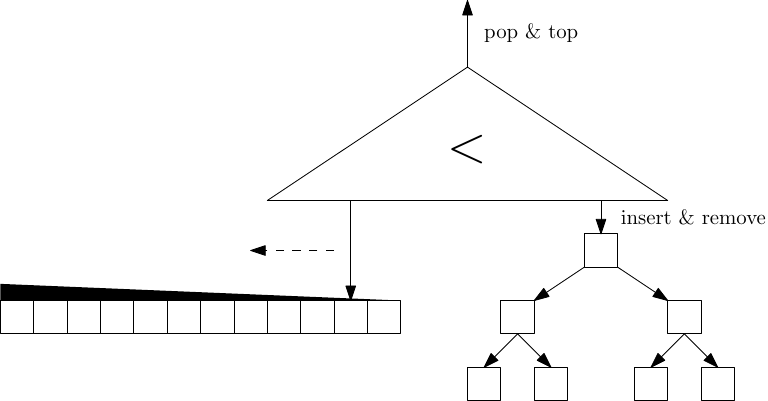}
		\caption{The priority queue consists of a \emph{static}, sorted array of input points 
			and a \emph{dynamic}, addressable PQ for intersection and deletion events.}
		\label{fig:prioq}
	\end{minipage}
\end{figure}

In order to support the required operations,
we use a doubly linked list of rays, with an AVL search tree on top  \cite{AVL}.
\figref{fig:sweepline_ds} shows a graphical representation of our data structure.
As the order of the rays along the sweepline is known at the time of insertion
-- see \alglnref{alg:gng}{insert} --
the \Ologn time search phase of a traditional AVL data structure can be omitted,
therefore insert operations can be performed in \Oone amortized time including rebalancing.
However, this optimization requires the need for parent pointers in the tree.
As always two neighboring rays are inserted into the sweepline data structure at the same time,
we implement a special insert operation for this case that only requires one rebalancing operation for both rays.
For removal operations, the position of the ray within the sweepline is known as well -- refer to \alglnref{alg:gng}{replace}.
Thus, similar to insert operations, no search phase is required for removals and the operation can be performed in \Oone amortized time.
The algorithm always removes the two neighboring rays $B_L$ and $B_R$ and and replaces them with $B^\ast$.
$B^\ast$ has the same left neighbor as $B_L$ and the same right neighbor as $B^R$.
Therefore, we can simply replace $B_L$ with $B^\ast$ in the data structure and just need to remove $B^R$,
leading to merely one rebalancing operation.

The search for the enclosing region of a point $p$ -- \alglnref{alg:gng}{find} --
is performed by finding the first ray $B_R$, currently  intersecting the sweepline, 
that has $p$ to its right.
The left neighbor $B_L$ of $B_R$, must have $p$ to its left or on it.
Therefore $B_L$ and $B_R$ enclose $p$ and $B_L^r = B_R^l = R_q$ gives the the region $p$ is contained in.
To facilitate searching, each internal node of the tree needs to refer to the rightmost ray in its subtree.
As rays are complex objects, we use pointers to the corresponding leaf to save memory.

\subsection{Priority Queue}\label{sec:prioq}

The priority queue $Q$ is initialized with all input points at the beginning of the algorithm.
During event processing, intersection and deletion events may be added and removed from $Q$,
therefore requiring an addressable priority queue.
The objects are ordered 
according to \eqnref{eq:proj},
thus keys are (exact) numerical values.
Our experiments show that, typically, for $n$ input points, only about \OnSqrt intersection and deletion events are in $Q$ at any given step.
Using the same PQ for all events would result in expensive dynamic PQ operations.
As input point events are static in $Q$,
we can use a two part data structure as shown in \figref{fig:prioq}.
Input point events are stored in an array 
-- sorted by priority in $Q$ --
with a pointer to the smallest unprocessed element.
Intersection and deletion events are stored in an actual addressable priority queue.
We use an addressable binary heap for this part of the data structure.
The \textsc{top} operation needs to compare the minimum element of the PQ
with the element pointed to in the array and return the minimum of both.
\textsc{Pop} either performs a regular pop on the PQ 
or moves the pointer of the array to the next larger element, accordingly.
\textsc{Insert} and \textsc{remove} operations can access the PQ directly,
as only this part of the data structure is dynamic.
Thus, the actual dynamic PQ is much smaller 
resulting in more efficient \textsc{siftDown}
and \textsc{siftUp} operations in the binary heap.
The smaller heap size not only reduces the tree height
but also makes the heap more cache friendly.

This optimization might be of interest for other algorithms 
that initialize their priority queue with all input points and only have a small number of dynamically added events
in their priority queue at any given time.
Note that the \emph{total} number of processed intersection and deletion events surpasses the number of input points by far,
however only a small number of these events are in the PQ at the same time.

\section{Evaluation}\label{sec:eval}

In this section we evaluate our implementation on a variety of datasets against other algorithms for Yao graph generation.

\subsection{Competing Algorithms}

\begin{figure}[tb]
	\centering
	\includegraphics[width=.3\textwidth]{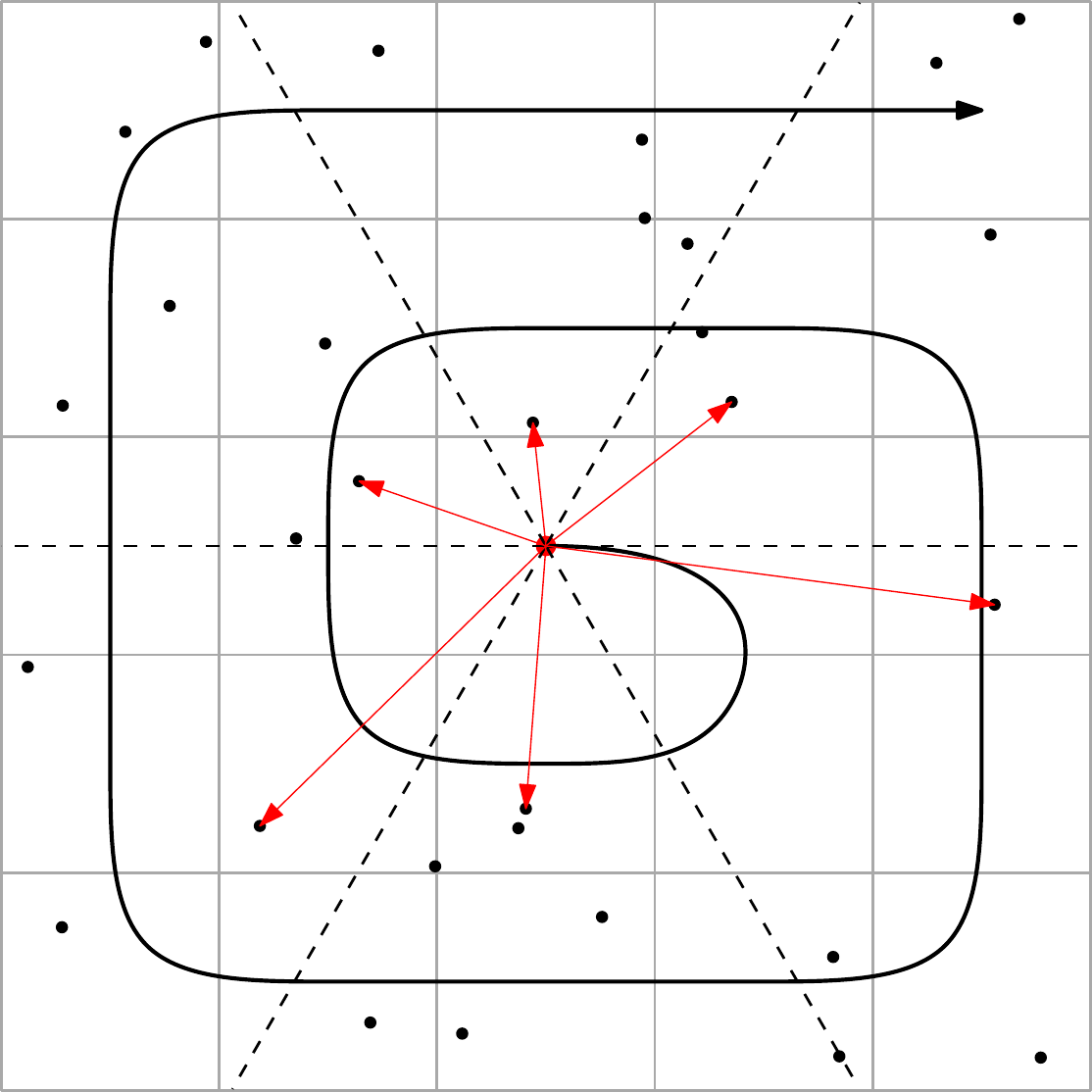}
	\caption{Grid-based Yao graph construction algorithm.
		The cone boundaries are represented by dashed lines.
		The algorithm visits grid cells in order of the thick curve.
		Found edges of the Yao graph are labeled in red.}
	\label{fig:grid_algo}
\end{figure}

As mentioned before, we are not aware of any previous implementations of Chang \etal's Yao graph algorithm.
Therefore, we evaluate our implementation against other algorithms to construct the Yao graph of a given point set.
Our main competitor is the Yao graph algorithm from the CGAL library's cone-based spanners package \cite{CGAL:Cone}.
As we are not aware of any other tuned implementations to construct Yao graphs,
we implement two other algorithms ourselves as competition.

First, we implement a \emph{naive} \OnSq algorithm that serves as a trivial baseline.
The algorithm compares the distance of all point pairs and determines the closest neighbor in each of the $k$ cones for a point.
It requires only two geometric predicates: distance comparison and oriented side of line test.
For a point pair $p$ and $q$, the cone $\mathcal{C}_i^p$ that $q$ lies in with respect to $p$ can first be approximated by 
$i =\nicefrac{\sphericalangle p q}{k}$.
Then, two oriented side of line tests suffice to exactly determine the cone $q$ lies in. 

Second, we implement a \emph{grid}-based algorithm.
The algorithm places all points in a uniform grid data structure \cite{Grid},
that splits the bounding box of all input points in \On equal sized cells.
For each input point $p$, the algorithm first visits $p$'s own grid cell and computes for each point $q$ in the cell
its distance to $p$ and the cone $q$ lies in with respect to $p$.
The algorithm then visits the grid cells surrounding $p$'s cell in a spiraling manner, refer to \figref{fig:grid_algo}.
For each visited cell, the algorithm computes the distances and cones for the points contained in it with respect to $p$ until all cones of point $p$ are settled.
A cone is settled, if a neighbor $v$ has been found within that cone
and no point in adjacent grid cells can be closer to $p$ than $v$.
Note that some cones may remain unsettled until all grid cells have been visited if no other input points lie within that cone for point $p$.
While the algorithm still has a \OnSq worst case time complexity, it performs much better in practice.

\subsection{Experimental Setup}
We test all algorithms on a variety of synthetic and real-world datasets.
We use input point sets distributed uniformly and normally in the unit square,
as well as points lying on the circumference of a circle and at the intersections of a grid \cite{Inputs}
-- the former being a worst case input for the grid algorithm, the latter being a bad case for numerical stability.
We furthermore use two real-world datasets:
intersections in road networks and star catalogs.
As road networks we use graphs from the $9^\text{th}$ DIMACS implementation challenge \cite{DIMACS}.
To generate a road network of a desired size $n$ from the \textsc{Full USA} graph, 
we use a random location and grow a rectangular area around it until at least $n$ nodes are within the area.
US cities feature many points on a grid  and therefore present a challenge for numerical stability.
We furthermore use the Gaia DR2 star catalog \cite{GaiaDR2},
which contains celestial positions for approximately \num{1.3} billion stars. 
We use a similar technique as for road networks to generate subgraphs of a desired size.
Here, we grow a cube around a random starting location until the desired number of stars fall within it.
We then project all stars onto the $xy$-plane as 2D input for our experiments.
\figref{fig:app:dist} in the appendix shows examples of our input datasets and resulting Yao graphs.

We implemented all algorithms in our C++ framework \textsc{YaoGraph}, available on Github.\footnote{%
\GitUrl}
Our code was compiled using GCC 12.1.0 with CGAL version 5.0.2.
All experiments were run on a server with an Intel Xeon Gold 6314U CPU with 32 cores and 512\,GiB of RAM.
For all input point distributions we used three different random seeds and, %
unless otherwise specified, we use $k = 6$ for all computations.

\subsection{Algorithmic Metrics}

\begin{figure}[tb]
	\centering
	\begin{subfigure}[t]{.5\textwidth}
		\centering
		\includegraphics[width=\textwidth]{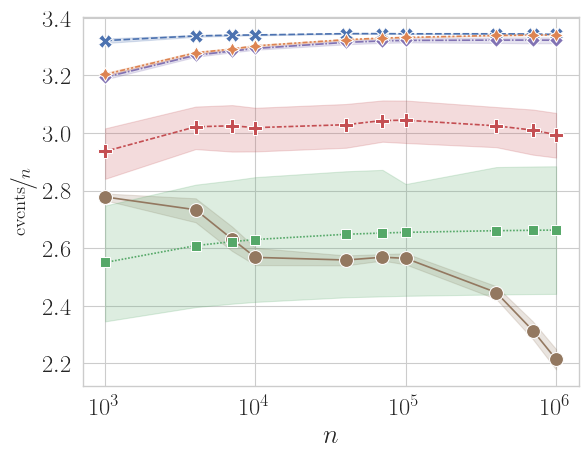}
		\caption{Total number of events processed.}
		\label{fig:eventsProcessed}
	\end{subfigure}\hfill
	\begin{subfigure}[t]{.5\textwidth}
		\centering
		\includegraphics[width=\textwidth]{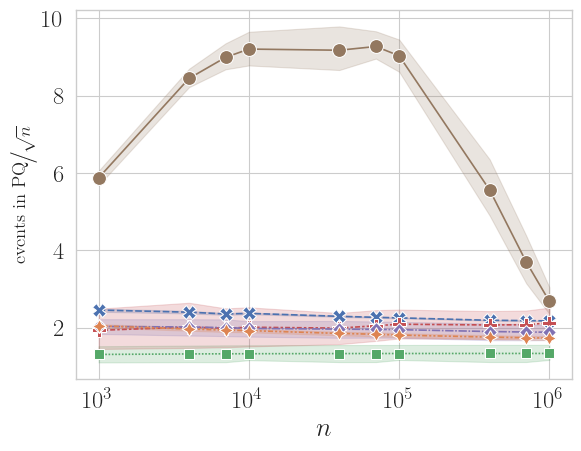}
		\caption{Maximum number of intersection and deletion events concurrently in PQ.}
		\label{fig:eventsInQueue}
	\end{subfigure}

	\begin{subfigure}[t]{.68\textwidth}
		\centering
		\includegraphics[width=\textwidth]{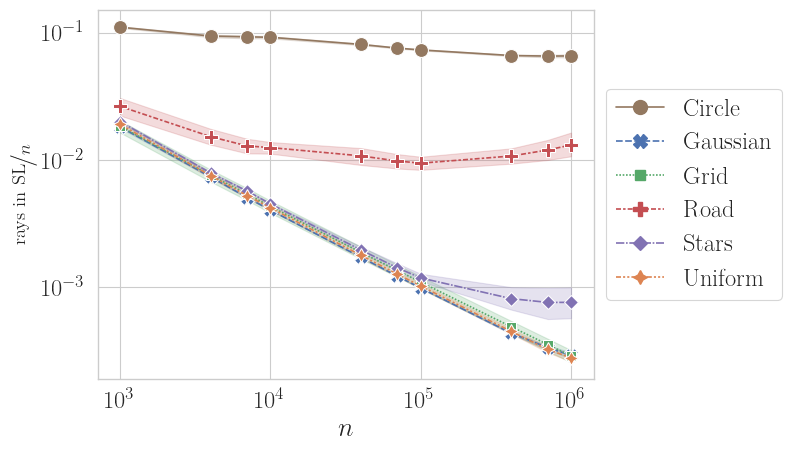}
		\caption{Maximum number of rays in sweepline data structure.}
		\label{fig:slSize}
	\end{subfigure}
	\caption{Statistics for varying input point distribution and point set sizes for $k=6$ cones.
		Error bands give the variation over the different cones being calculated.}
	\label{fig:algStats}
\end{figure}

Firstly, we discuss relevant properties of the sweepline algorithm.
\figref{fig:eventsProcessed} shows the number of events processed per input point by the algorithm.
Each input point has one input point event and generates \num{2.3} intersection and/or deletion events on average,
with very little variance with regard to input size and distribution -- except for the grid distribution and road graphs.
Both exhibit larger variance, depending on whether the cone's boundaries coincide with grid lines or not.
\figref{fig:eventsInQueue} shows the maximum number of intersection and deletion events that are in the priority queue at any given time during the algorithm execution.
This number scales with \OnSqrt for most studied inputs,
which motivates our choice of two part priority queue as discussed in \secref{sec:prioq}.
The behavior of the circle distribution requires further investigation.	
The maximum number of rays in the sweepline data structure at any point during algorithm execution shows no clear scaling behavior, refer to \figref{fig:slSize}.
It scales with \OnSqrt for most synthetic input sets,
but approaches a constant fraction of the input size for the circle (\qty{\approx 10}{\percent}), road (\qty{\approx 1}{\percent}) and star (\qty{\approx 0.1}{\percent}) datasets.

\subsection{Runtime Evaluation}

\begin{figure}[tb]
	\centering
	\begin{subfigure}[t]{.5\textwidth}
		\centering
		\includegraphics[width=\textwidth]{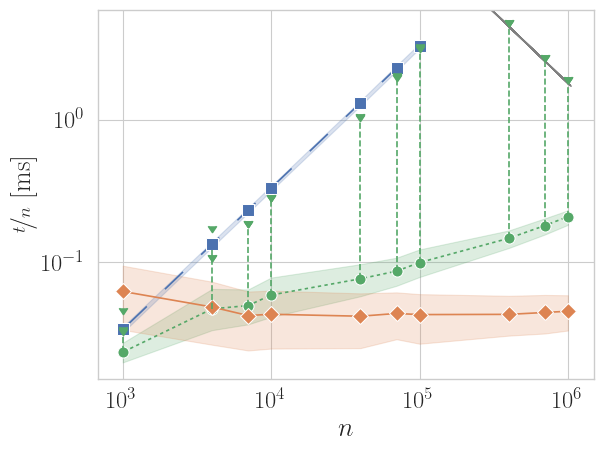}
		\caption{Inexact kernel.}
		\label{fig:runtimeInexact}
	\end{subfigure}\hfill
	\begin{subfigure}[t]{0.5\textwidth}
		\centering
		\includegraphics[width=\textwidth]{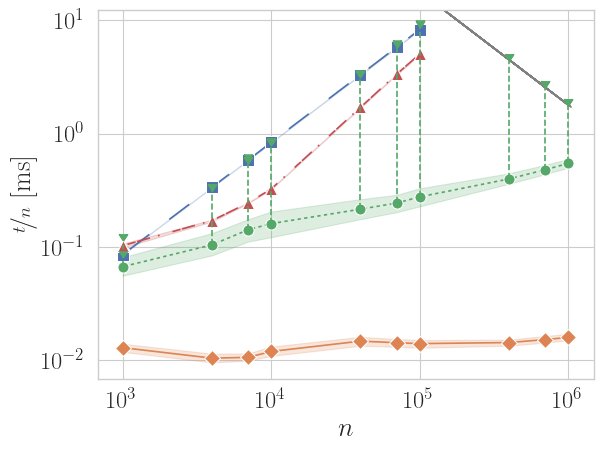}
		\caption{CGAL EPIC kernel.}
		\label{fig:runtimeCGALInexact}
	\end{subfigure}\hfill
	\begin{subfigure}[t]{0.67\textwidth}
		\centering
		\includegraphics[width=\textwidth]{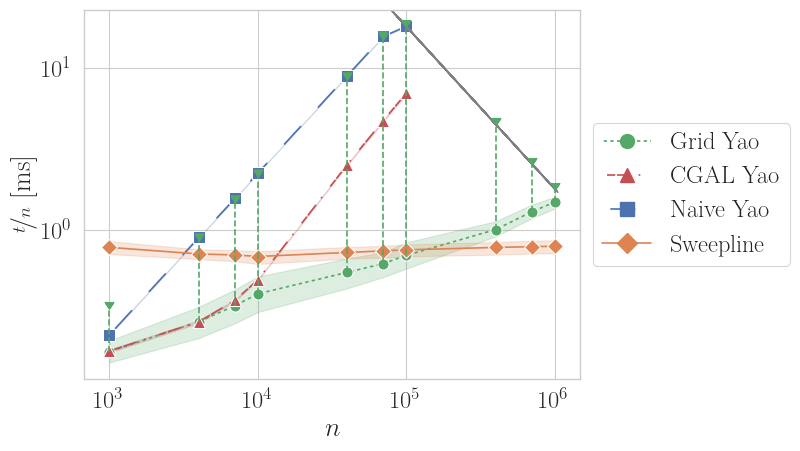}
		\caption{CGAL EPEC kernel.}
		\label{fig:runtimeCGALExact}		
	\end{subfigure}\hfill
	\caption{Algorithm runtime experiments.
		Experiments over varying input sizes are performed with $k = 6$ cones.  
		Error bands give the runtime variation over the different input point distributions.
		For the grid algorithm, the circle distribution is plotted separately with triangular markers.
		The gray line represents the time limit of \qty{30}{\minute} per algorithm.
		Experiments over varying number of cones are with $n=\num{1e5}$ uniformly distributed input points.}
	\label{fig:algRuntime}
\end{figure}
	
\begin{figure}[tb]
	\centering
	\includegraphics[width=.69\textwidth]{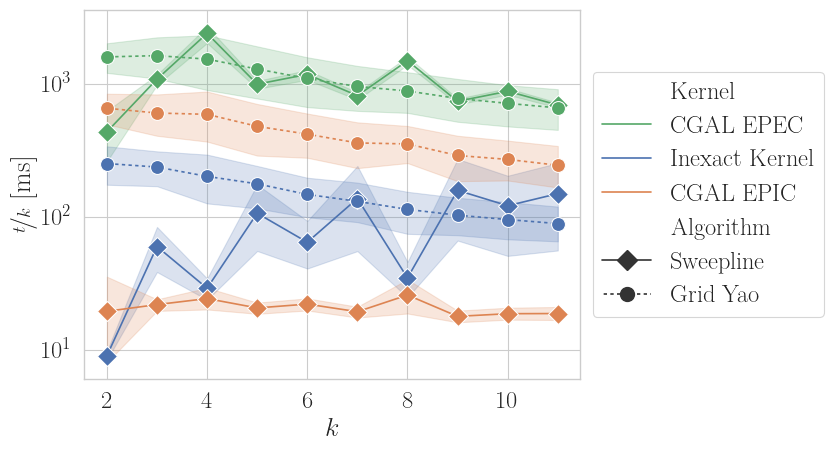}
	\caption{Algorithm runtime experiments.
		 	Experiments over varying number of cones are with $n=\num{1e5}$ uniformly distributed input points.}
	\label{fig:algRuntimeCones}
\end{figure}

\figref{fig:algRuntime} shows the results of our runtime experiments.
Plots \figref{fig:runtimeInexact} to \figref{fig:runtimeCGALExact}
show the (scaled) running time of the algorithms, displaying variations due to input distributions as error bands.
\figref{fig:app:algRuntime} in the appendix gives a more detailed picture of the runtime for the different distributions.
Note that only the grid and the sweepline algorithm are sensitive to the input point distribution.
As previously seen in \figref{fig:eventsProcessed}, the number of processed events by the sweepline algorithm is relatively constant for all distributions.
Therefore only little variation is seen in the runtime of the algorithm.
This also shows, that the size of the sweepline data structure has only negligible influence on the algorithm runtime,
as no higher runtime is observed for the road or circle datasets.
Our inexact kernel shows more runtime variation than CGAL's highly optimized kernels,
mainly due to the grid distribution with its many points directly on cone boundaries.
The sweepline algorithm clearly outperforms CGAL's Yao graph implementation.
Furthermore, even though it requires much more involved computations, it is superior to the simple grid algorithm for non-exact constructions.
Only for exact constructions, large inputs are required to negate the more expensive operations of the sweepline algorithm.
The exact construction kernel leads to runtime overhead of \num{100} compared to the EPIC kernel.
However, if points lie directly on cone boundaries, 
exact constructions are necessary to obtain correct results, as seen in \figref{fig:app:dist:grid} in the appendix.
The data dependency is  more pronounced for the grid algorithm, 
which performs well for most datasets but degenerates to the naive algorithm for the circle distribution,
due to the many empty grid cells in the circle's interior.

To compute a Yao graph with $k$ cones, the sweepline algorithm requires $k$ passes.
This linear relationship can be seen in \figref{fig:algRuntimeCones}.
The grid algorithm has no dependency on $k$ -- except for the size of the neighborhood of a point.
However, our experiments show, that the runtime of the algorithm increases with increasing $k$.
We attribute this to the fact, that more grid cells need to be visited in order to settle all cones of a point $p$,
since with narrower cones, 
chances are higher that no points lying in a specific cone of $p$ are within a visited grid cell.
We did not perform these experiments with the naive algorithm or the CGAL algorithm, 
due to their  long runtimes.
CGAL's algorithm also requires one pass per cone,
whereas the naive algorithm's runtime does not depend on $k$.

\section{Conclusion}\label{sec:outro}

We present the -- to the best of our knowledge -- first implementation of Chang \etal's optimal \Onlogn time Yao graph algorithm.
Our implementation uses carefully engineered data structures and algorithmic operations and
thereby outperforms current publicly available Yao graph implementations -- particularly CGAL's cone-based spanners package -- by at least an order of magnitude.
We furthermore present a very simple grid-based Yao graph algorithm that also outperforms CGAL's implementation,
but is inferior to Chang \etal's algorithm for larger input.
However, the algorithm could be further improved by using a precomputed mapping of the grid neighborhood to cones,
in order to only visit grid cells that can contain points in hitherto unsettled cones.
Moreover, the algorithm is trivially parallelizable over the input points, 
whereas Chang \etal's algorithm can only be easily parallelized over the $k$ cones.
The parallelization within one sweepline pass remains for future work.

\FloatBarrier
\newpage

\bibliographystyle{plainurl}
\bibliography{yao}

\begin{thebibliography}{10}

\bibitem{AVL}
Georgy~Maksimovich Adelson-Velsky and Evgeny~Mikhailovich Landis.
\newblock An algorithm for organization of information.
\newblock In {\em Doklady Akademii Nauk}, volume 146, pages 263--266. Russian
  Academy of Sciences, 1962.

\bibitem{Grid}
V.~Akman, W.R. Franklin, M.~Kankanhalli, and C.~Narayanaswami.
\newblock Geometric computing and uniform grid technique.
\newblock {\em Computer-Aided Design}, 21(7):410--420, 1989.
\newblock \href {https://doi.org/https://doi.org/10.1016/0010-4485(89)90125-5}
  {\path{doi:https://doi.org/10.1016/0010-4485(89)90125-5}}.

\bibitem{BenOtt79}
J.~L. Bentley and T.~A. Ottmann.
\newblock Algorithms for reporting and counting geometric intersections.
\newblock {\em IEEE Transactions on Computers}, pages 643--647, 1979.

\bibitem{CGAL:Kernel}
Herv{\'e} Br{\"o}nnimann, Andreas Fabri, Geert-Jan Giezeman, Susan Hert,
  Michael Hoffmann, Lutz Kettner, Sylvain Pion, and Stefan Schirra.
\newblock {2D} and {3D} linear geometry kernel.
\newblock In {\em {CGAL} User and Reference Manual}. {CGAL Editorial Board},
  {5.5.1} edition, 2022.
\newblock URL:
  \url{https://doc.cgal.org/5.5.1/Manual/packages.html#PkgKernel23}.

\bibitem{YaoOpt}
Maw~Shang Chang, Nen-Fu Huang, and Chuan-Yi Tang.
\newblock An optimal algorithm for constructing oriented voronoi diagrams and
  geographic neighborhood graphs.
\newblock {\em Information Processing Letters}, 35(5):255--260, 1990.
\newblock URL:
  \url{https://www.sciencedirect.com/science/article/pii/0020019090900542},
  \href {https://doi.org/https://doi.org/10.1016/0020-0190(90)90054-2}
  {\path{doi:https://doi.org/10.1016/0020-0190(90)90054-2}}.

\bibitem{GaiaDR2}
Gaia Collaboration.
\newblock Gaia data release 2. summary of the contents and survey properties.
\newblock {\em arXiv}, (abs/1804.09365), 2018.

\bibitem{DIMACS}
Camil Demetrescu, Andrew~V Goldberg, and David~S Johnson.
\newblock {\em The shortest path problem: Ninth DIMACS implementation
  challenge}, volume~74.
\newblock American Mathematical Soc., 2009.

\bibitem{FortuneSL}
Steven Fortune.
\newblock A sweepline algorithm for voronoi diagrams.
\newblock {\em Algorithmica}, 2(1):153--174, 1987.

\bibitem{Inputs}
Daniel Funke, Peter Sanders, and Vincent Winkler.
\newblock Load-balancing for parallel delaunay triangulations.
\newblock In Ramin Yahyapour, editor, {\em Euro-Par 2019: Parallel Processing},
  pages 156--169, Cham, 2019. Springer International Publishing.

\bibitem{ThetaSL}
Giri Narasimhan and Michiel Smid.
\newblock {\em Geometric spanner networks}.
\newblock Cambridge University Press, 2007.

\bibitem{YaoWireless}
Christian Schindelhauer, Klaus Volbert, and Martin Ziegler.
\newblock Geometric spanners with applications in wireless networks.
\newblock {\em Computational Geometry}, 36(3):197--214, 2007.

\bibitem{CGAL:Cone}
Weisheng Si, Quincy Tse, and Fr{\'e}d{\'e}rik Paradis.
\newblock Cone-based spanners.
\newblock In {\em {CGAL} User and Reference Manual}. {CGAL Editorial Board},
  {5.4} edition, 2022.
\newblock URL:
  \url{https://doc.cgal.org/5.4/Manual/packages.html#PkgConeSpanners2}.

\bibitem{YaoMST}
Andrew Chi-Chih Yao.
\newblock On constructing minimum spanning trees in k-dimensional spaces and
  related problems.
\newblock {\em SIAM Journal on Computing}, 11(4):721--736, 1982.
\newblock \href {https://arxiv.org/abs/https://doi.org/10.1137/0211059}
  {\path{arXiv:https://doi.org/10.1137/0211059}}, \href
  {https://doi.org/10.1137/0211059} {\path{doi:10.1137/0211059}}.

\bibitem{YaoLocalWireless}
Xiujuan Zhang, Jiguo Yu, Wei Li, Xiuzhen Cheng, Dongxiao Yu, and Feng Zhao.
\newblock Localized algorithms for yao graph-based spanner construction in
  wireless networks under sinr.
\newblock {\em IEEE/ACM Transactions on Networking}, 25(4):2459--2472, 2017.
\newblock \href {https://doi.org/10.1109/TNET.2017.2688484}
  {\path{doi:10.1109/TNET.2017.2688484}}.

\end{thebibliography}

\newpage
\FloatBarrier
\appendix

\section{Evaluation}

\FloatBarrier
\subsection{Input Point Distributions}

\newlength{\distfigwidth}
\setlength{\distfigwidth}{.25\textwidth}
\begin{figure}[tbh]
	\centering
	\captionsetup[subfigure]{justification=centering}
	\begin{subfigure}{\textwidth}
	\centering
	\begin{subfigure}[t]{\distfigwidth}
		\centering
		\includegraphics[width=\textwidth]{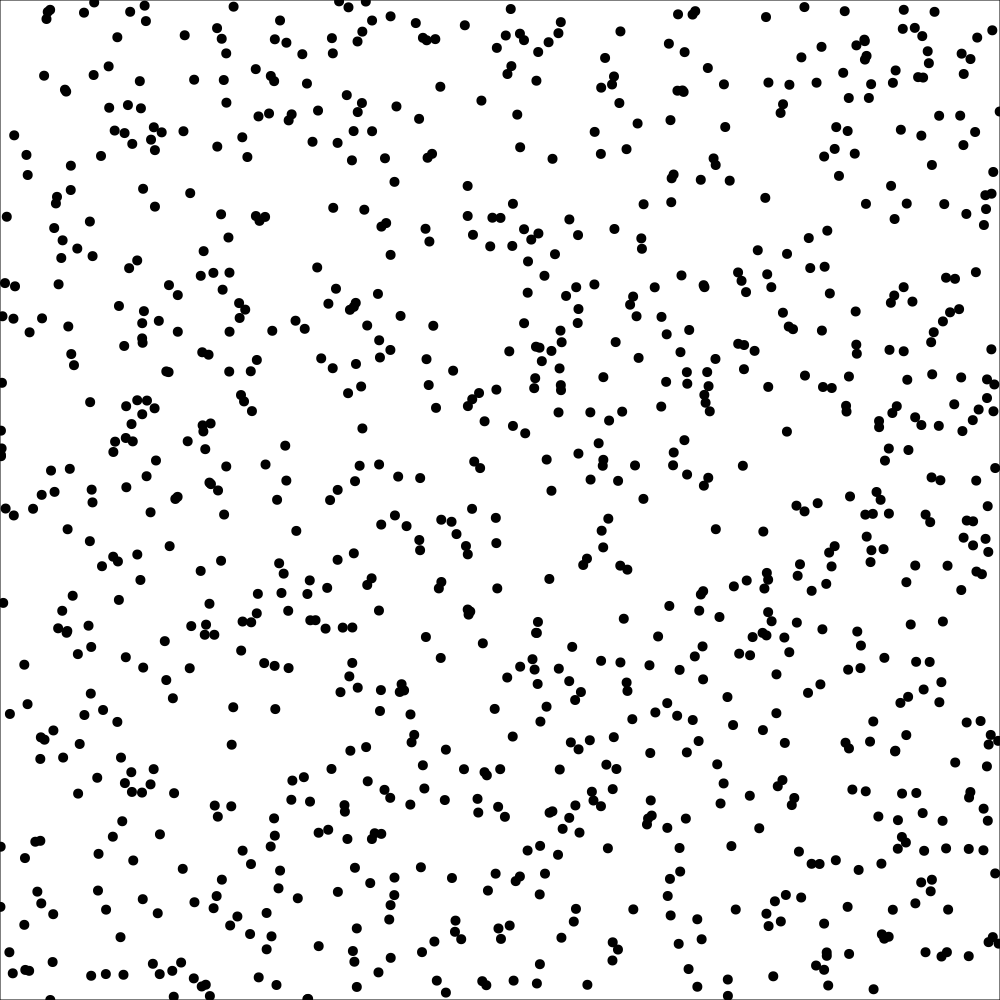}
	\end{subfigure}\hspace{.1\textwidth}
	\begin{subfigure}[t]{\distfigwidth}
		\centering
		\includegraphics[width=\textwidth]{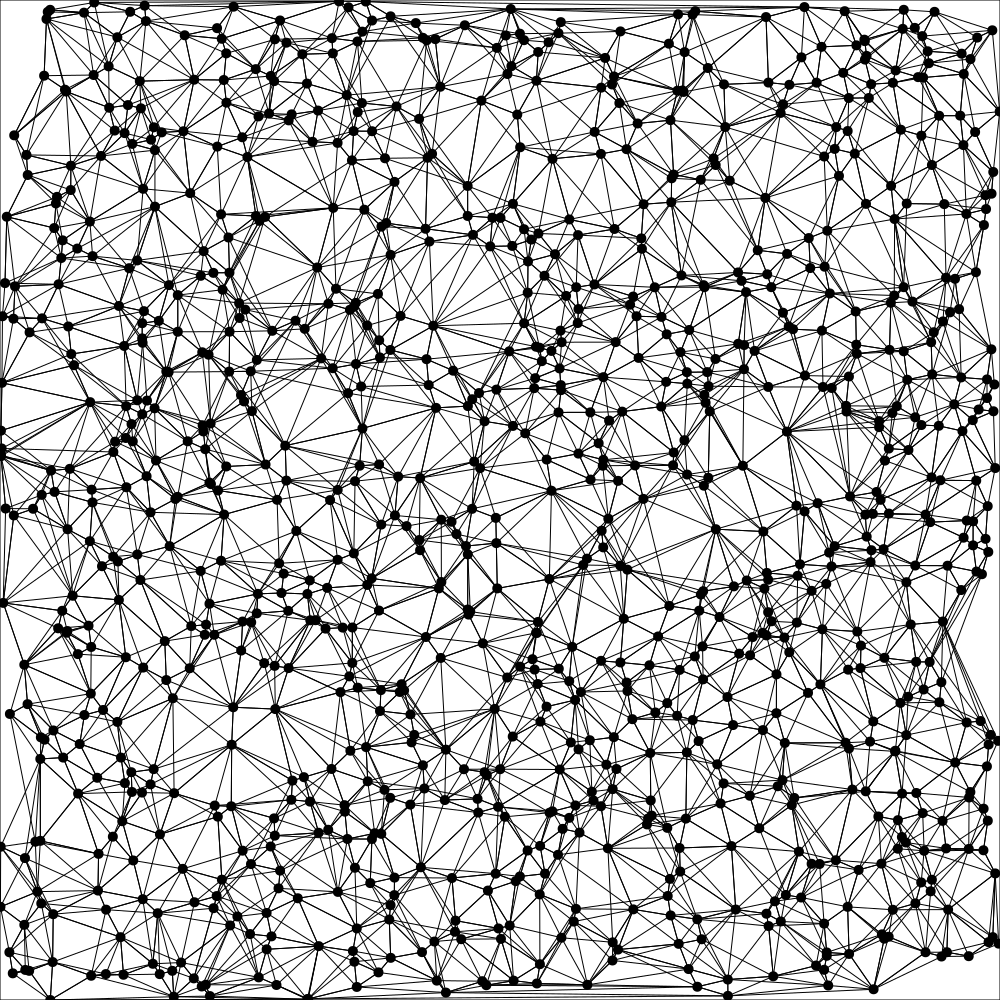}
	\end{subfigure}
	\caption{Uniform distribution.}
	\end{subfigure}

	\begin{subfigure}{\textwidth}
	\centering
	\begin{subfigure}[t]{\distfigwidth}
		\centering
		\includegraphics[width=\textwidth]{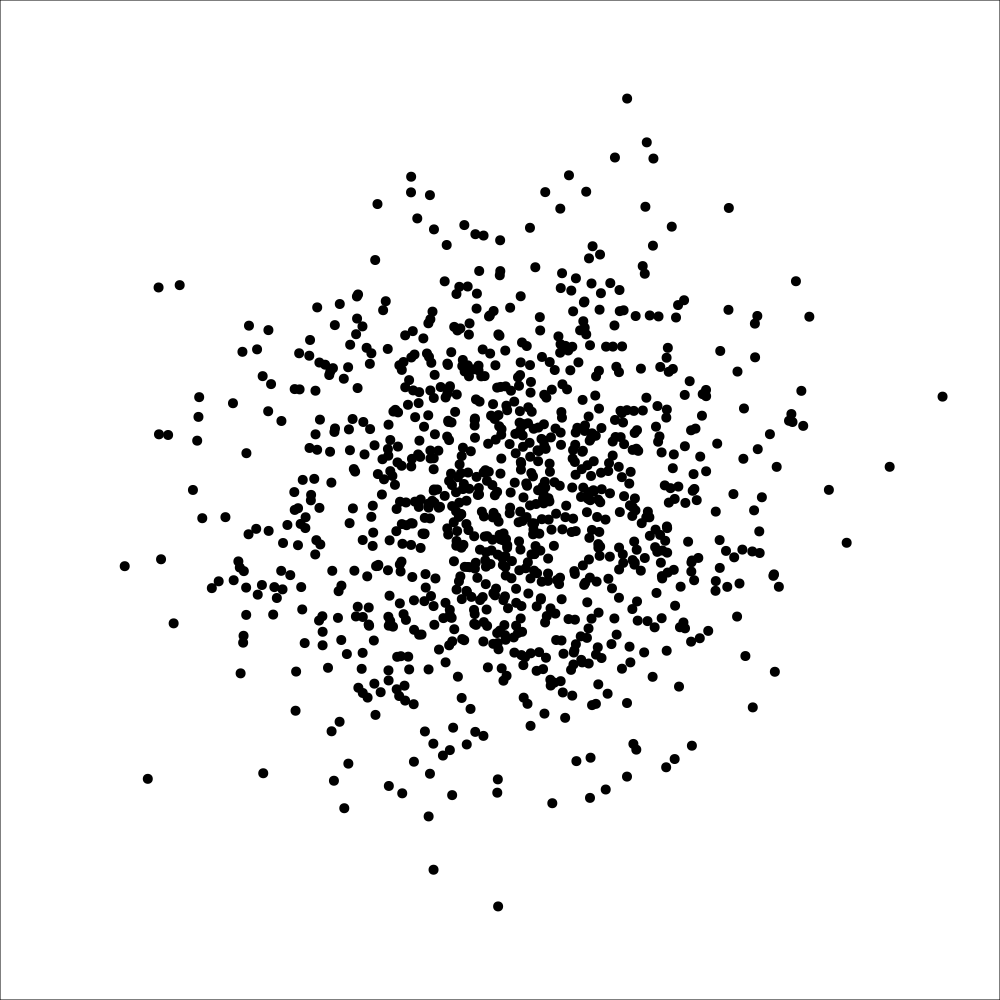}
	\end{subfigure}\hspace{.1\textwidth}
	\begin{subfigure}[t]{\distfigwidth}
		\centering
		\includegraphics[width=\textwidth]{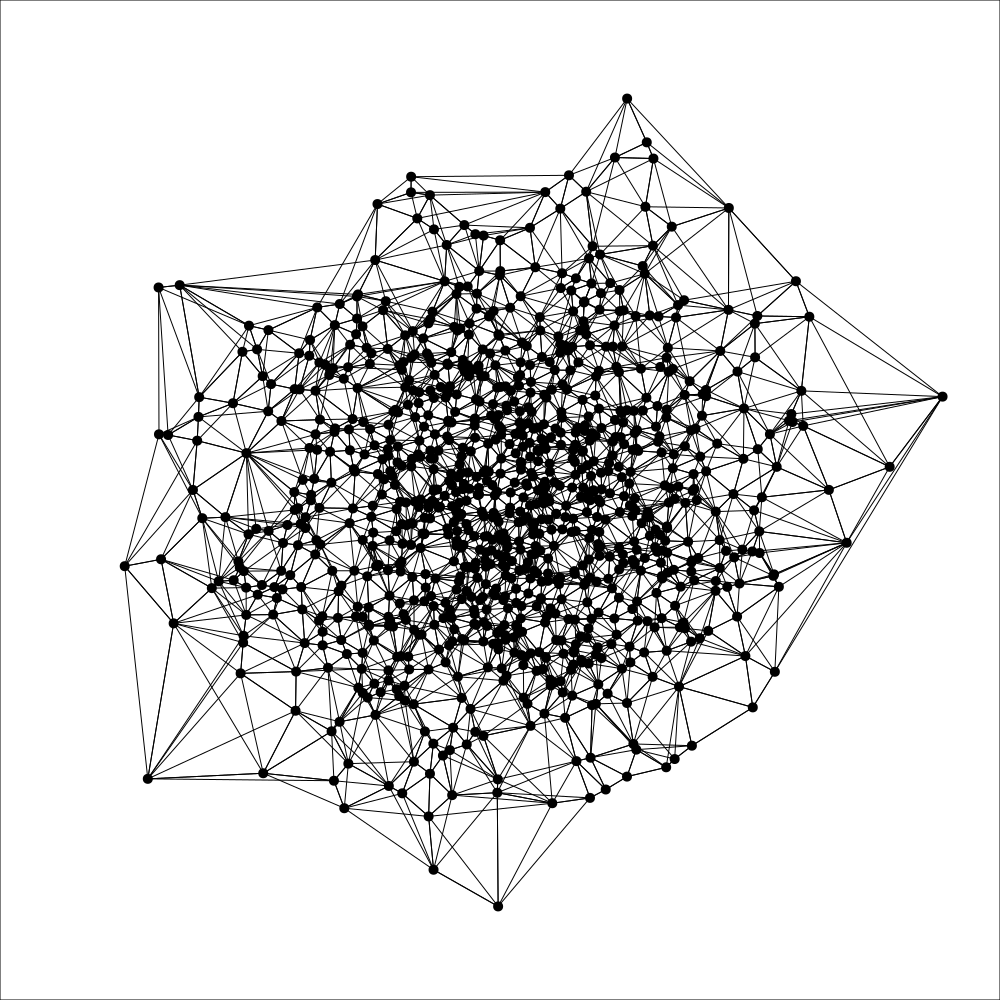}
	\end{subfigure}
	\caption{Gaussian distribution.}
	\end{subfigure}
	\begin{subfigure}{\textwidth}
	\centering
	\begin{subfigure}[t]{\distfigwidth}
		\centering
		\includegraphics[width=\textwidth]{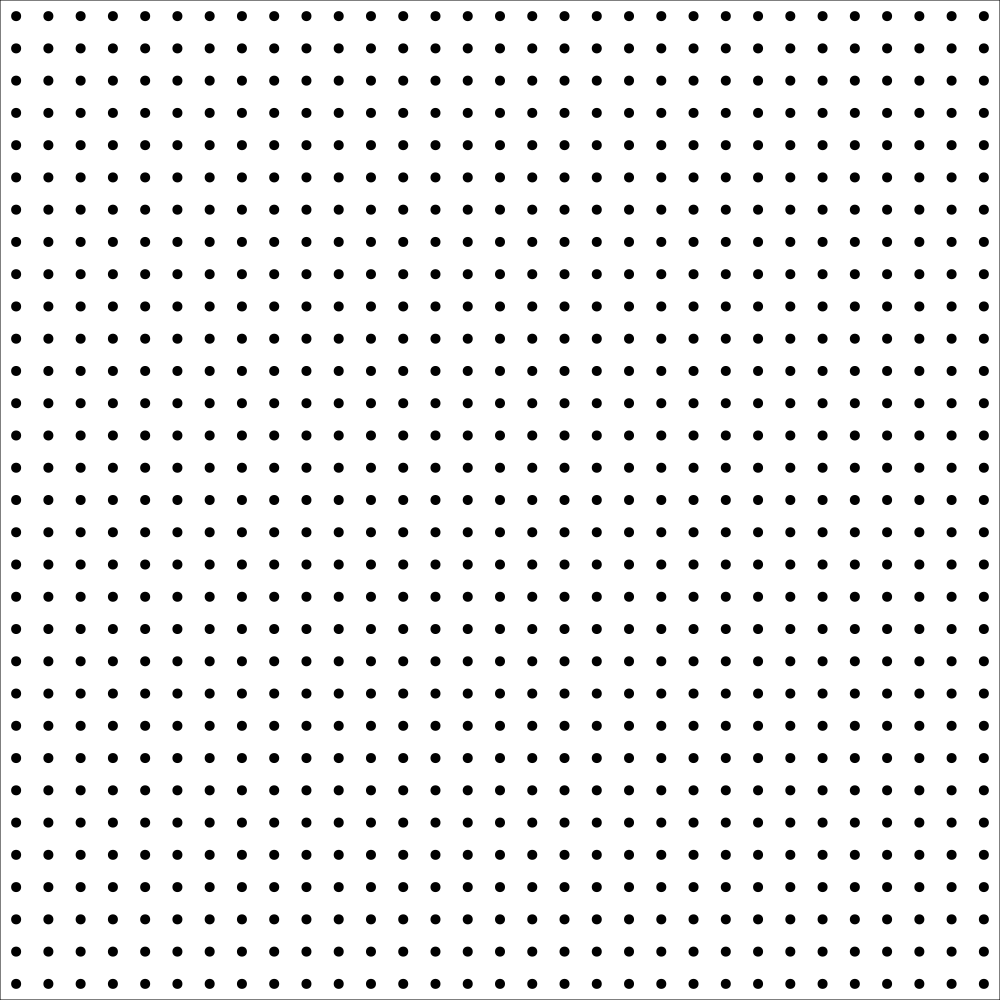}
	\end{subfigure}\hspace{.1\textwidth}
	\begin{subfigure}[t]{\distfigwidth}
		\centering
		\includegraphics[width=\textwidth]{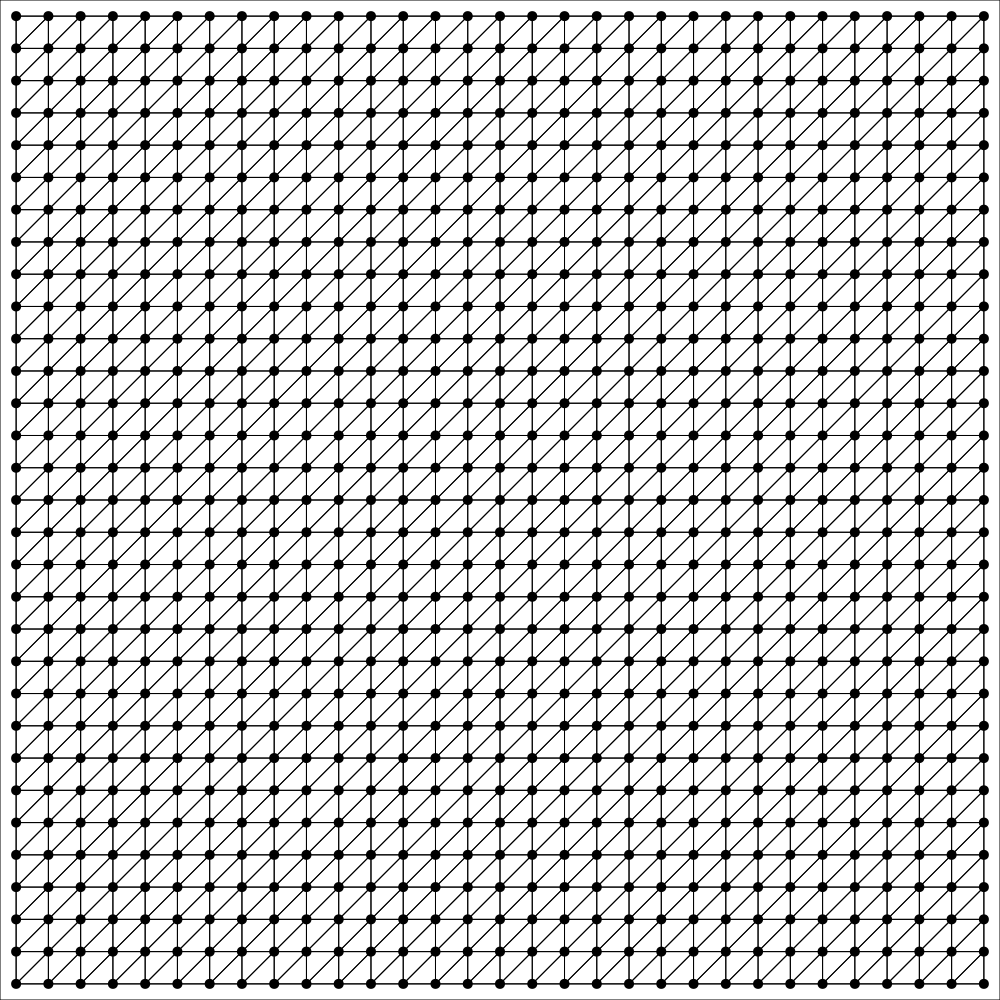}
		EPEC kernel
	\end{subfigure}\hspace{.1\textwidth}
	\begin{subfigure}[t]{\distfigwidth}
		\centering
		\includegraphics[width=\textwidth]{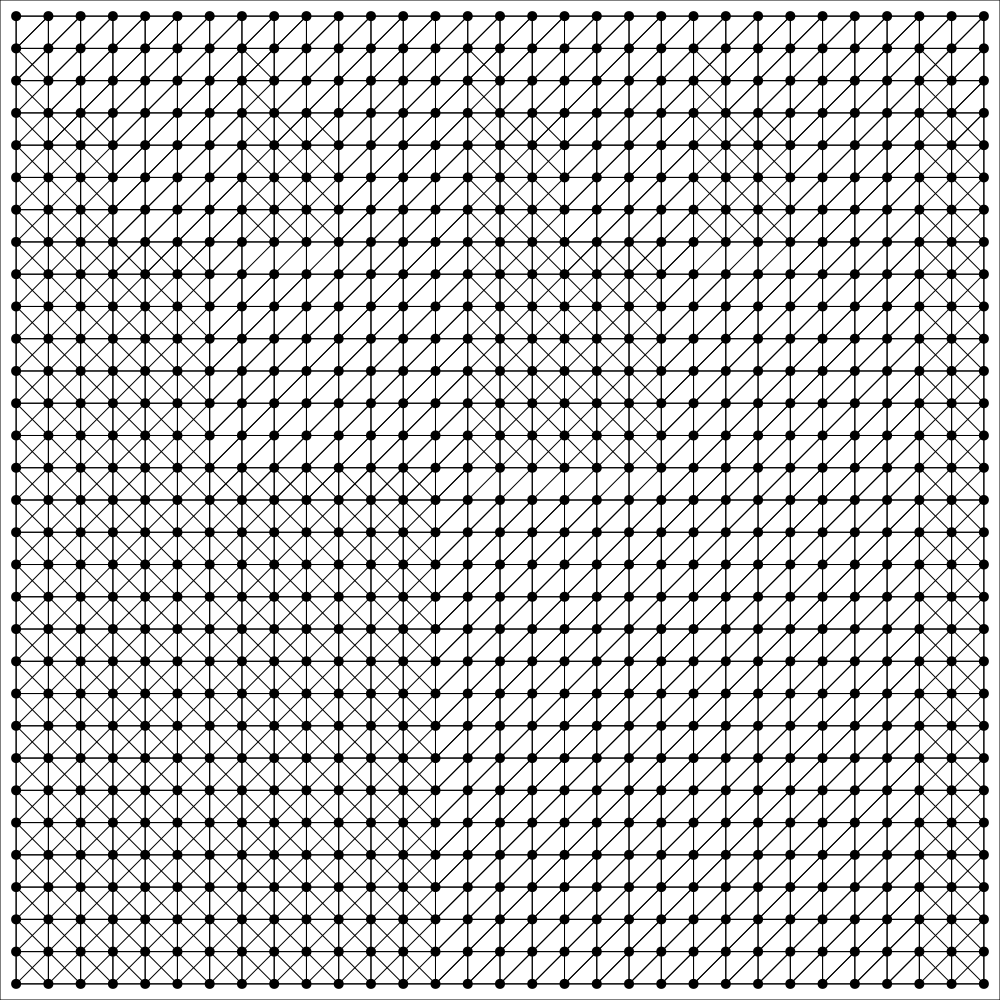}
		EPIC kernel
	\end{subfigure}
	\caption{Grid distribution.}
	\label{fig:app:dist:grid}
	\end{subfigure}
\end{figure}

\begin{figure}[tbh]\ContinuedFloat
	\centering
	\captionsetup[subfigure]{justification=centering}

	\begin{subfigure}{\textwidth}
	\centering
	\begin{subfigure}[t]{\distfigwidth}
		\centering
		\includegraphics[width=\textwidth]{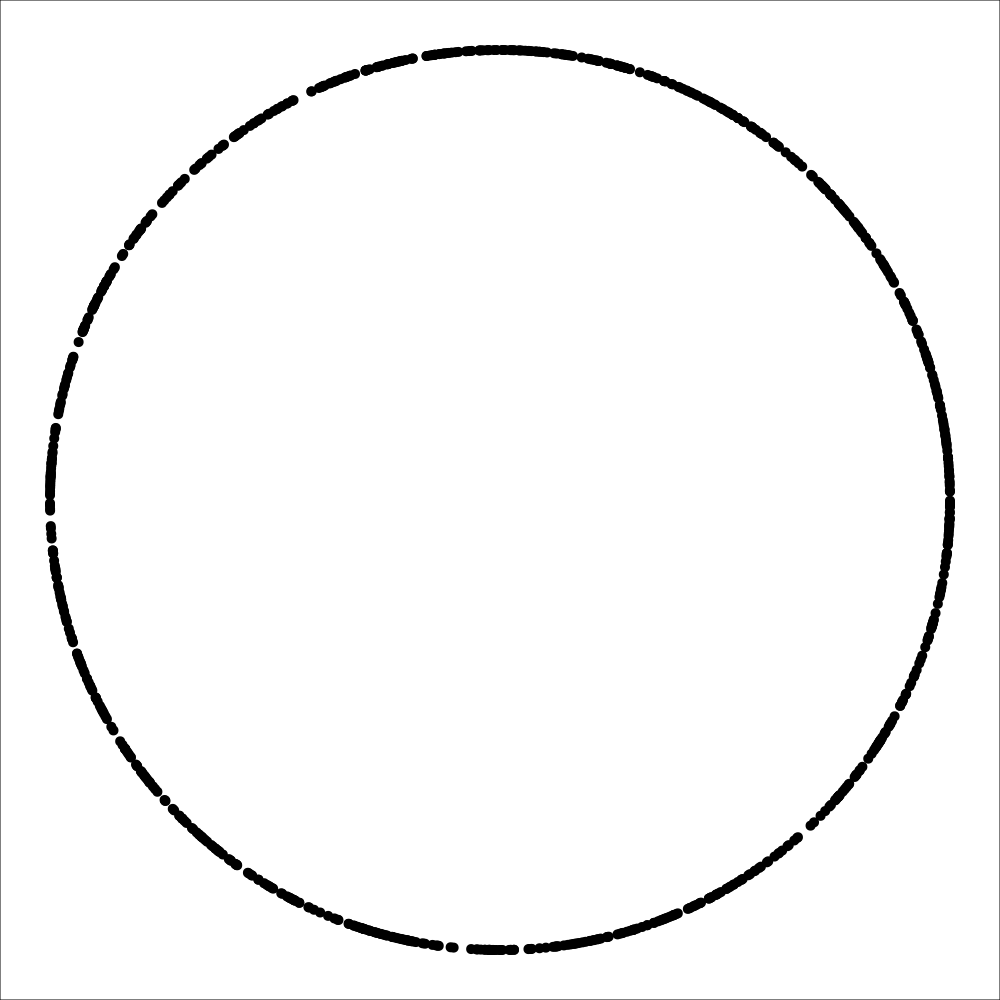}
	\end{subfigure}\hspace{.1\textwidth}
	\begin{subfigure}[t]{\distfigwidth}
		\centering
		\includegraphics[width=\textwidth]{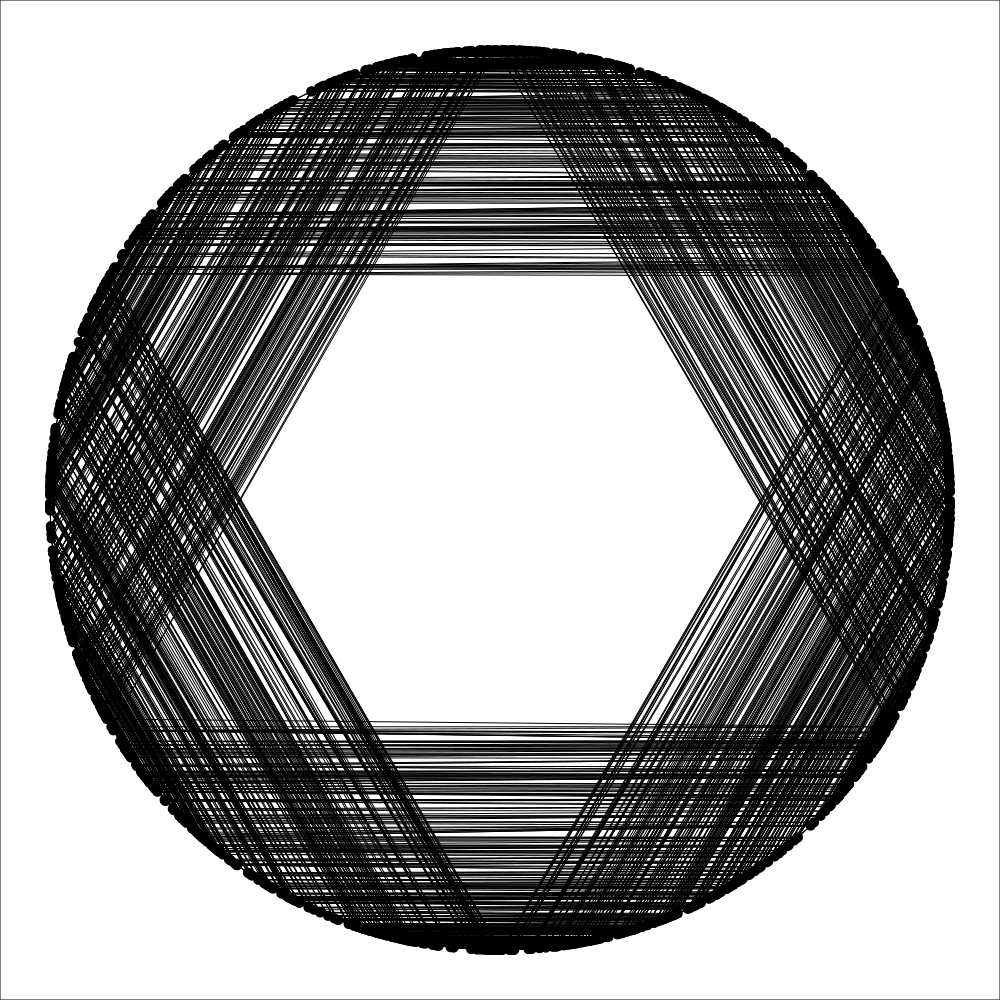}
	\end{subfigure}
	\caption{Circle distribution.}
	\end{subfigure}

	\begin{subfigure}{\textwidth}
	\centering
	\begin{subfigure}[t]{\distfigwidth}
		\centering
		\includegraphics[width=\textwidth]{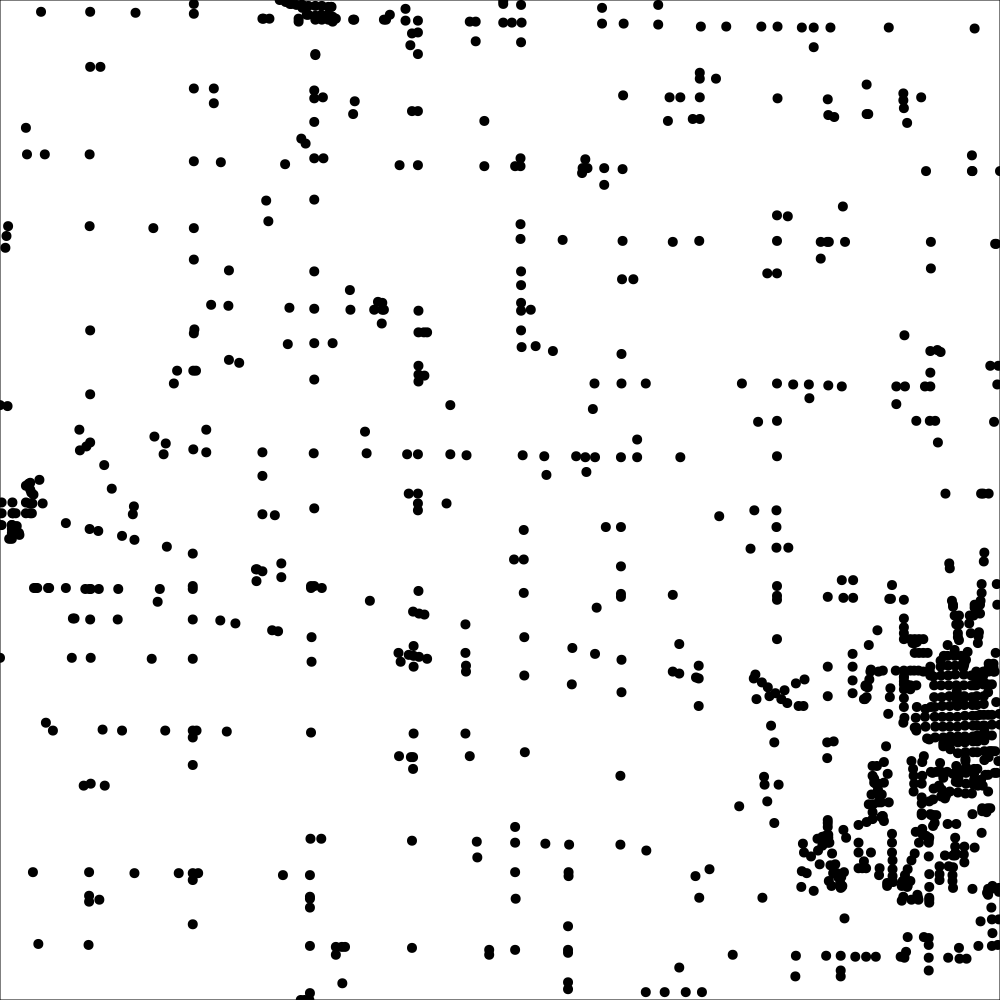}
	\end{subfigure}\hspace{.1\textwidth}
	\begin{subfigure}[t]{\distfigwidth}
		\centering
		\includegraphics[width=\textwidth]{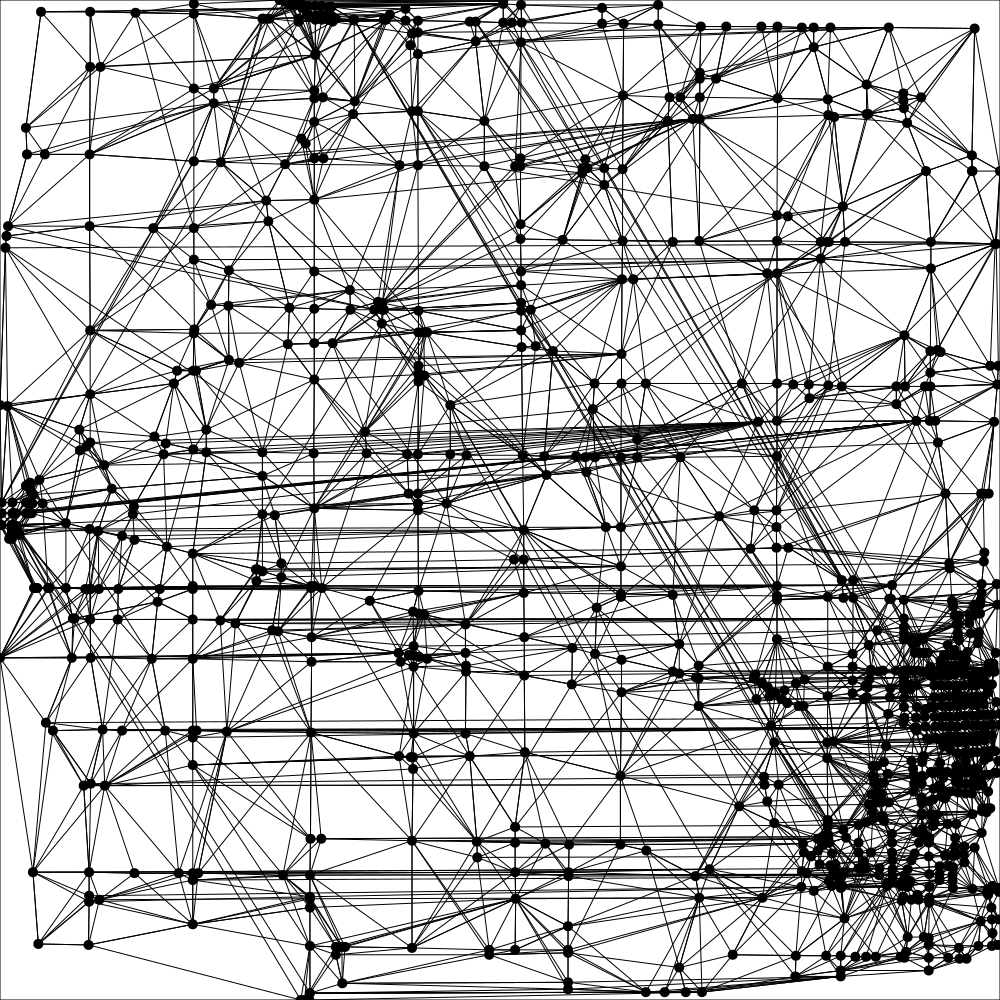}
	\end{subfigure}
	\caption{Road dataset.}
	\end{subfigure}

	\begin{subfigure}{\textwidth}
	\centering
	\begin{subfigure}[t]{\distfigwidth}
		\centering
		\includegraphics[width=\textwidth]{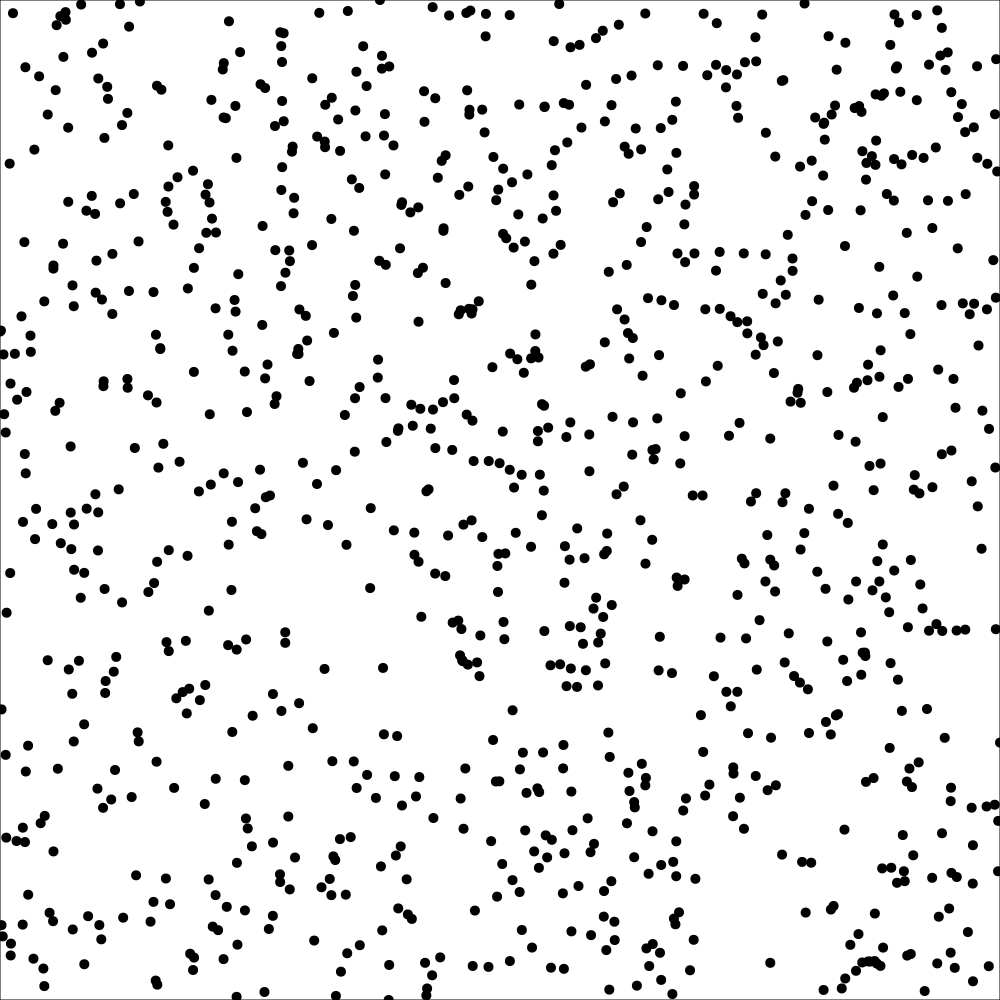}
	\end{subfigure}\hspace{.1\textwidth}
	\begin{subfigure}[t]{\distfigwidth}
		\centering
		\includegraphics[width=\textwidth]{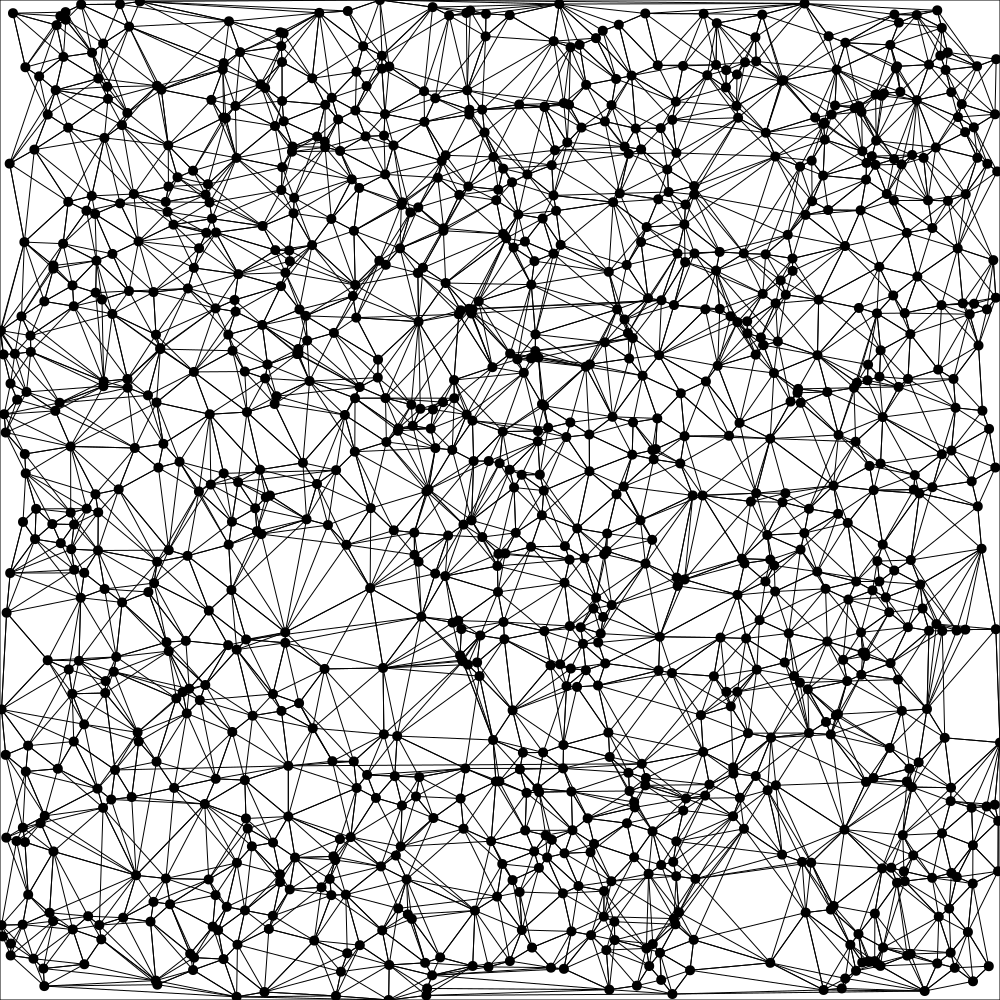}
	\end{subfigure}
	\caption{Star dataset.}
	\end{subfigure}

	\caption{Input distributions for $n = \num{1000}$ points and resulting Yao graph for $k = 6$.
	For the grid distribution, the resulting graphs from exact constructions and inexact constructions are shown.}
	\label{fig:app:dist}
\end{figure}

\FloatBarrier
\clearpage
\subsection{Example Execution}

\newlength{\execfigwidth}
\setlength{\execfigwidth}{.24\textwidth}
\begin{figure}[tbh]
	\centering
	\begin{subfigure}{\execfigwidth}
	\centering
	\includegraphics[width=\textwidth]{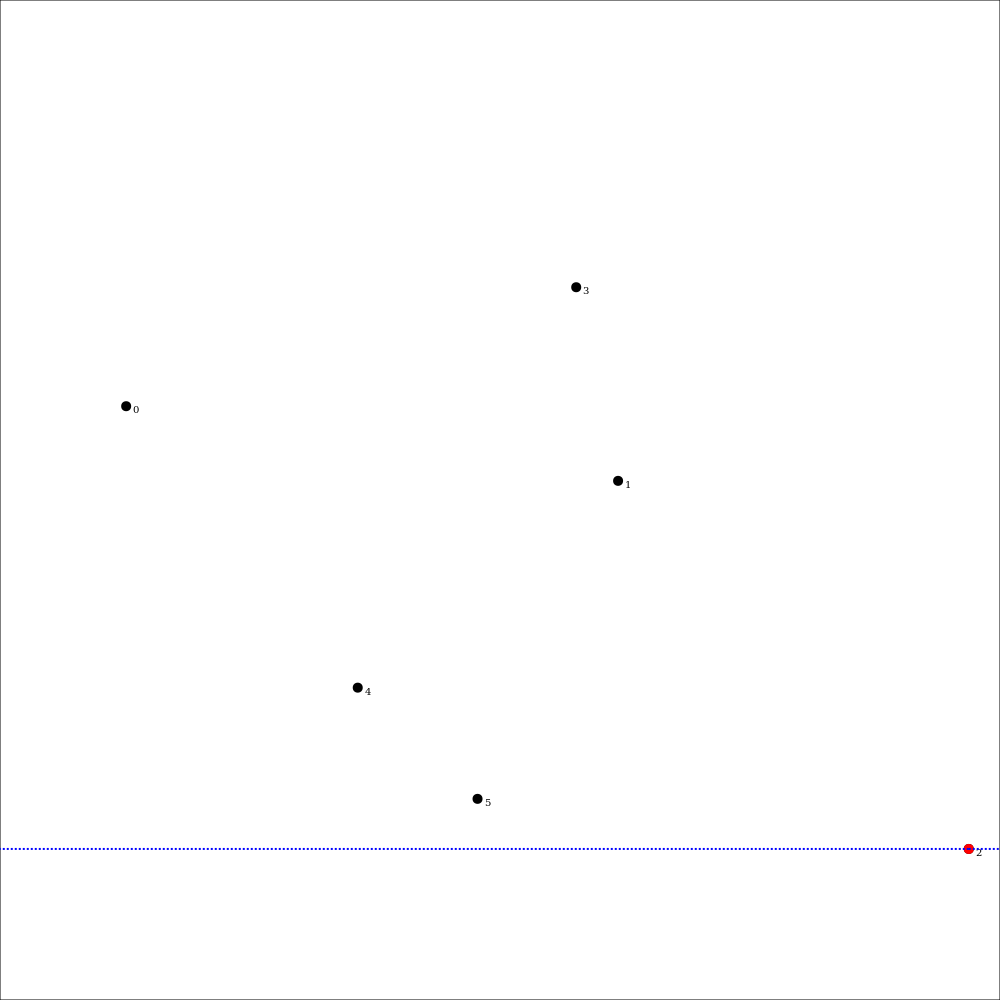}
	\end{subfigure}
	\begin{subfigure}{\execfigwidth}
	\centering
	\includegraphics[width=\textwidth]{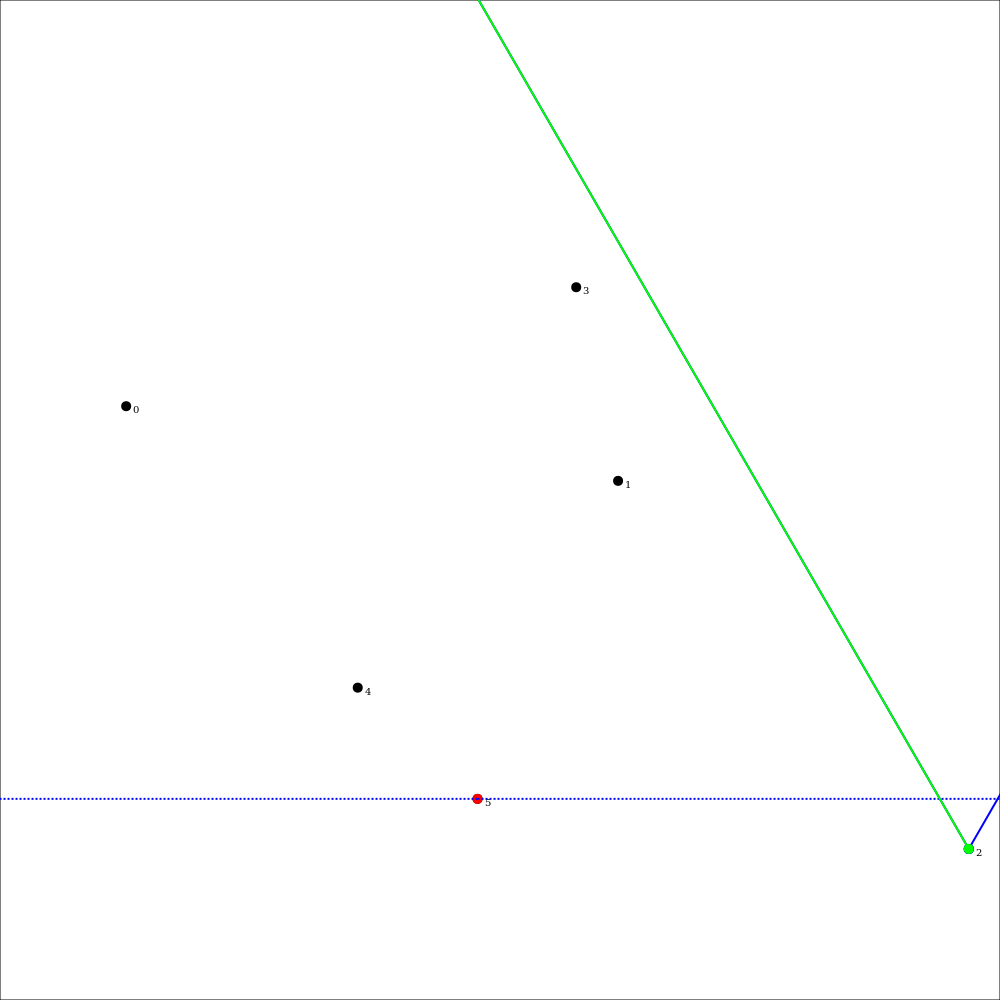}
	\end{subfigure}
	\begin{subfigure}{\execfigwidth}
	\centering
	\includegraphics[width=\textwidth]{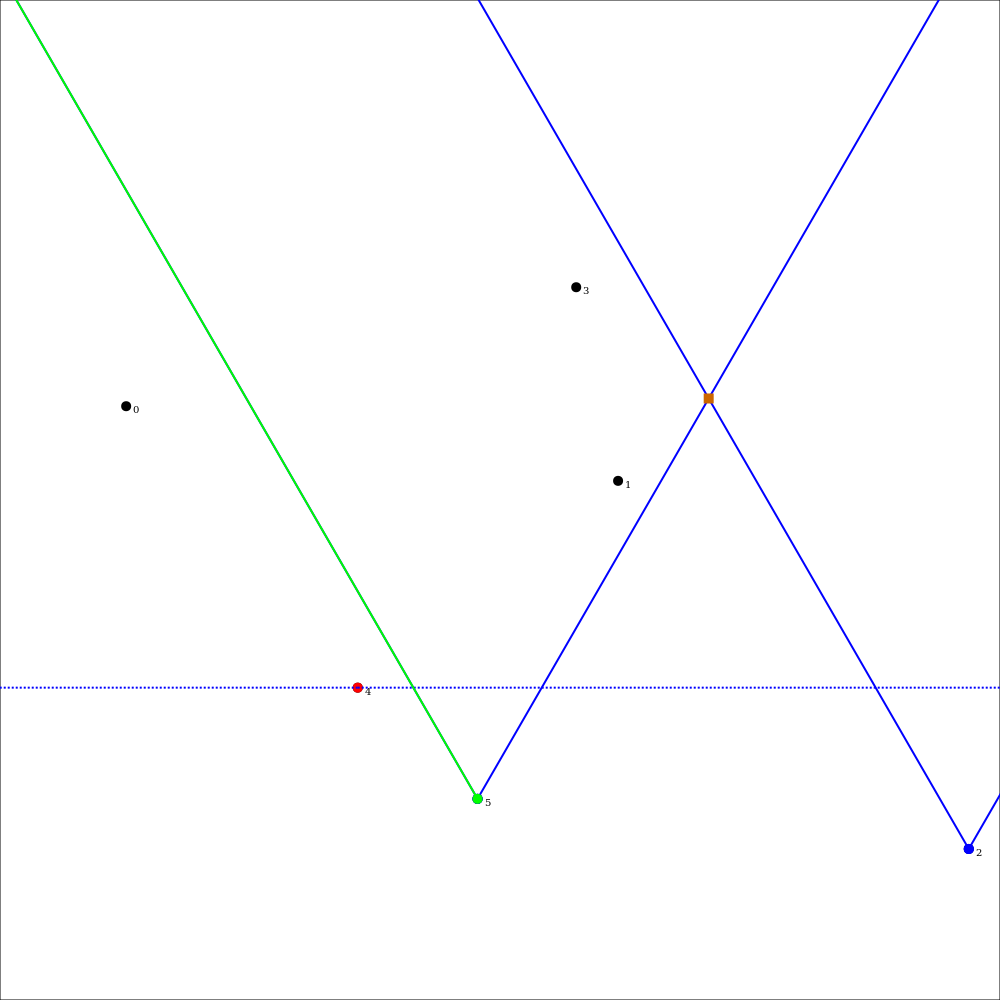}
	\end{subfigure}
	\begin{subfigure}{\execfigwidth}
	\centering
	\includegraphics[width=\textwidth]{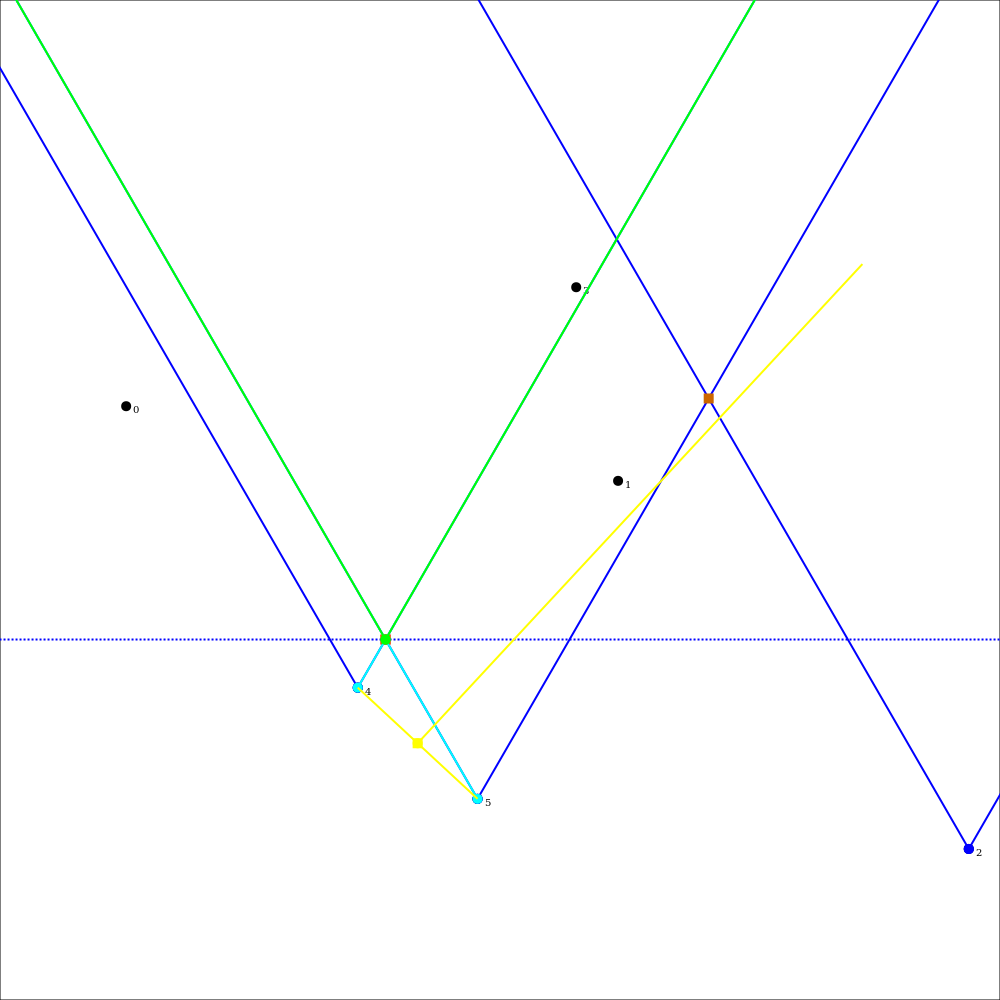}
	\end{subfigure}
	\begin{subfigure}{\execfigwidth}
	\centering
	\includegraphics[width=\textwidth]{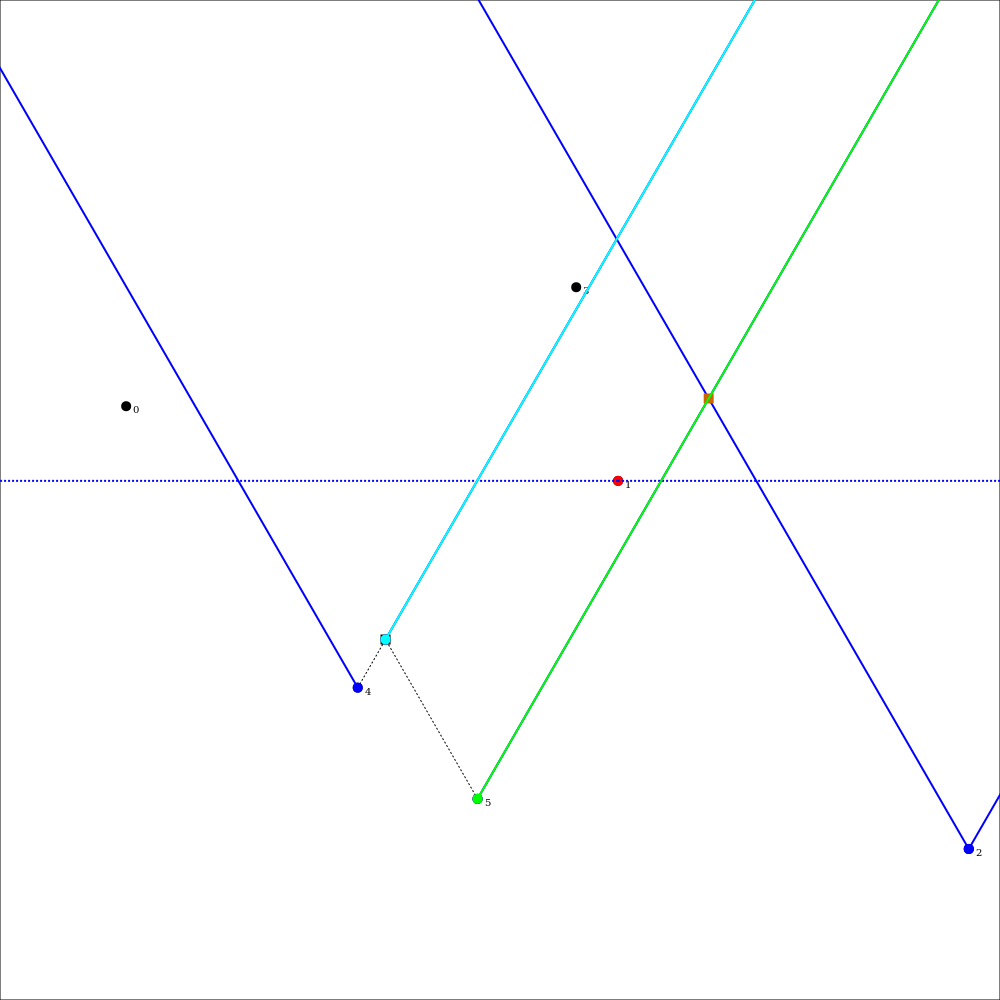}
	\end{subfigure}
	\begin{subfigure}{\execfigwidth}
	\centering
	\includegraphics[width=\textwidth]{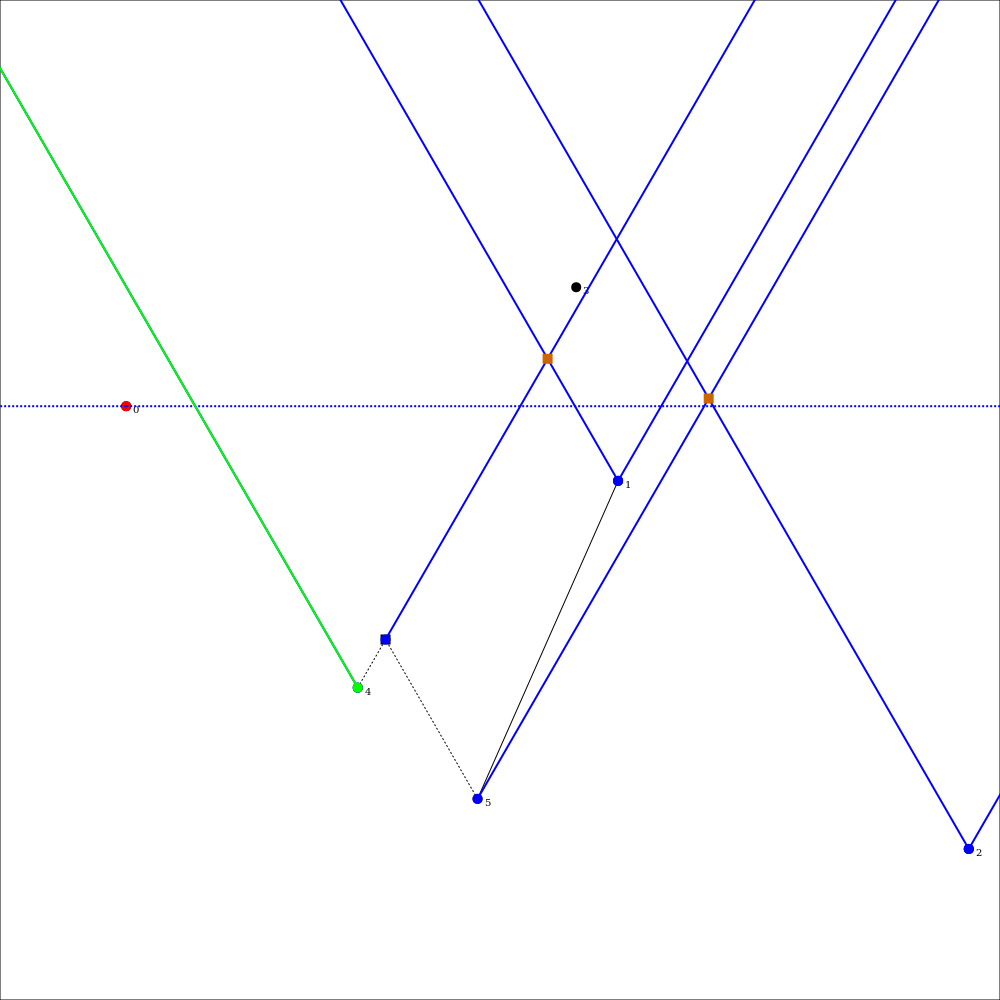}
	\end{subfigure}
	\begin{subfigure}{\execfigwidth}
	\centering
	\includegraphics[width=\textwidth]{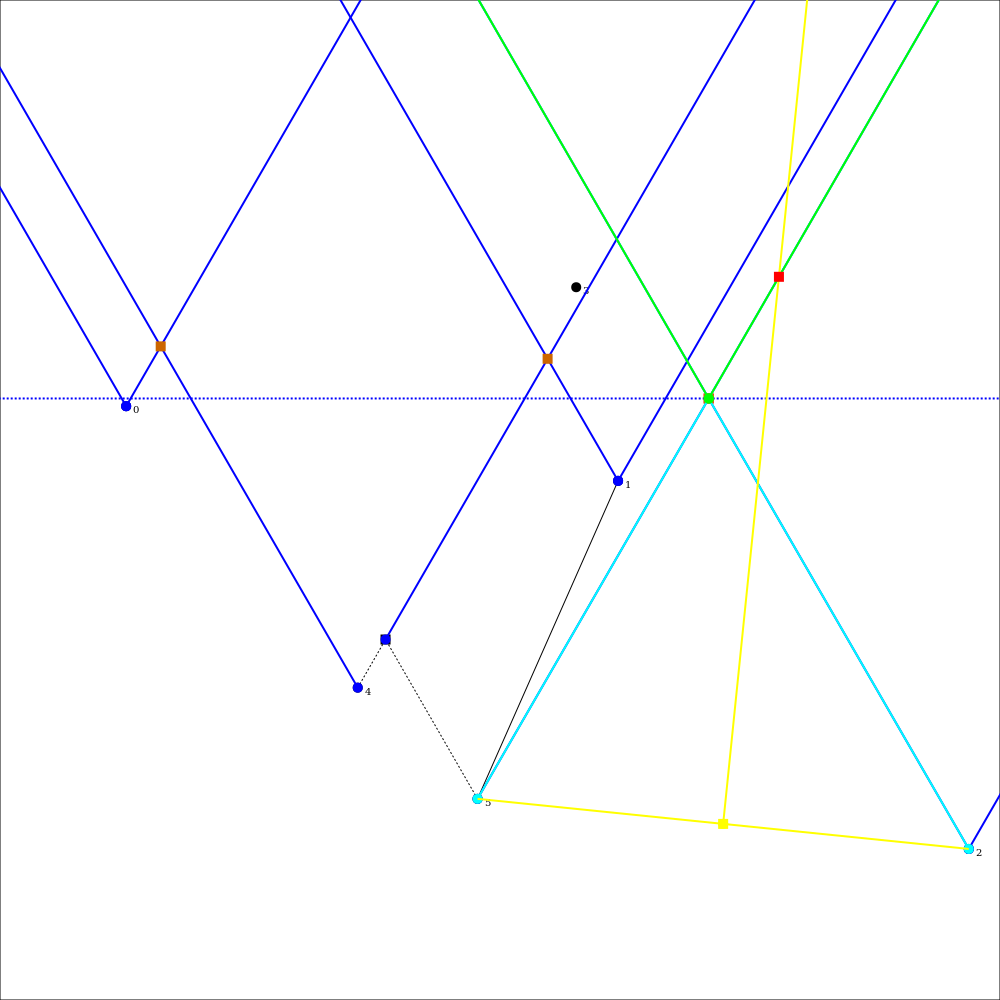}
	\end{subfigure}
	\begin{subfigure}{\execfigwidth}
	\centering
	\includegraphics[width=\textwidth]{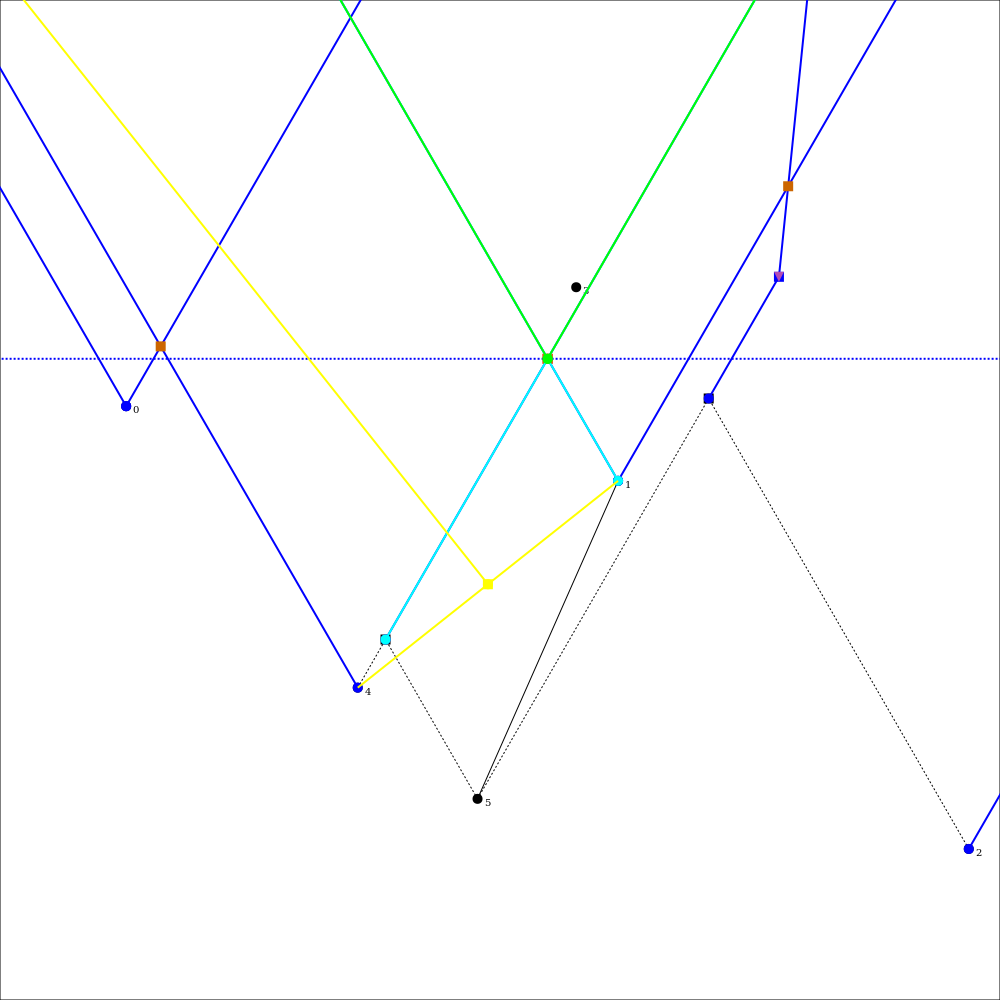}
	\end{subfigure}
	\begin{subfigure}{\execfigwidth}
	\centering
	\includegraphics[width=\textwidth]{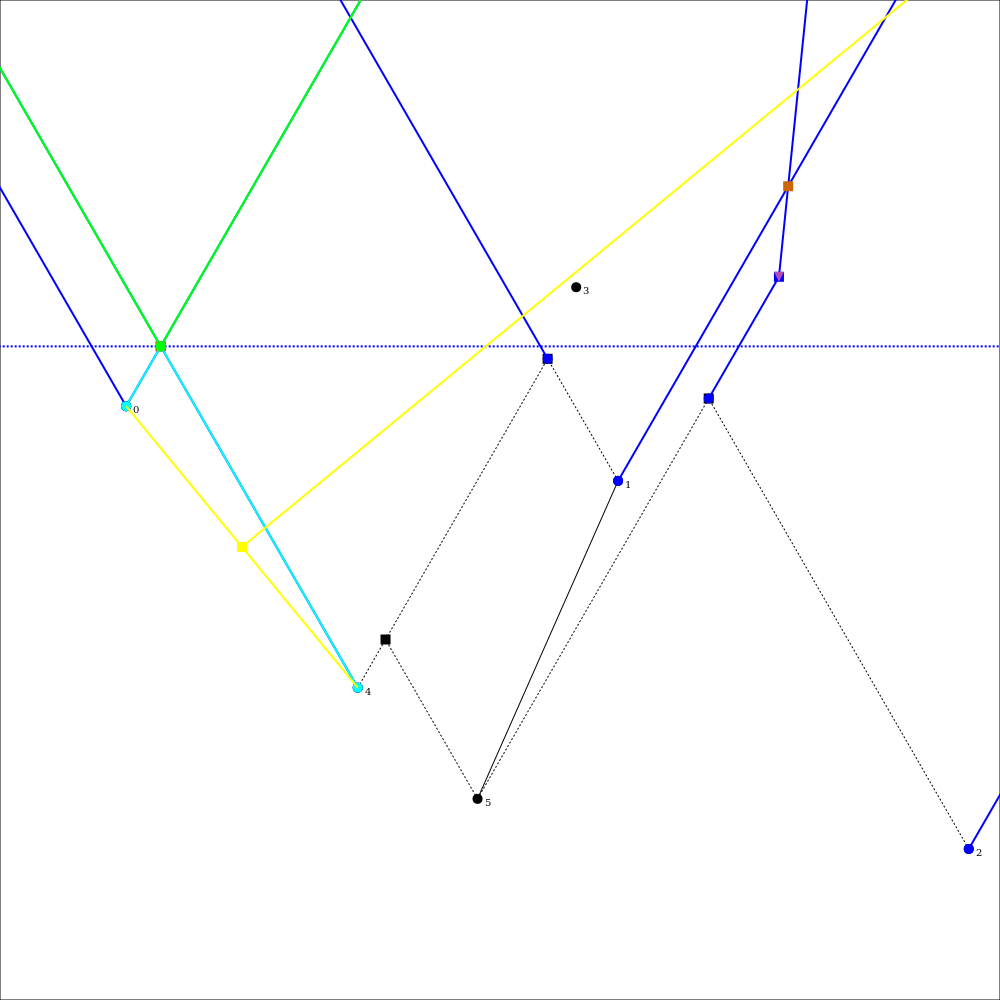}
	\end{subfigure}
	\begin{subfigure}{\execfigwidth}
	\centering
	\includegraphics[width=\textwidth]{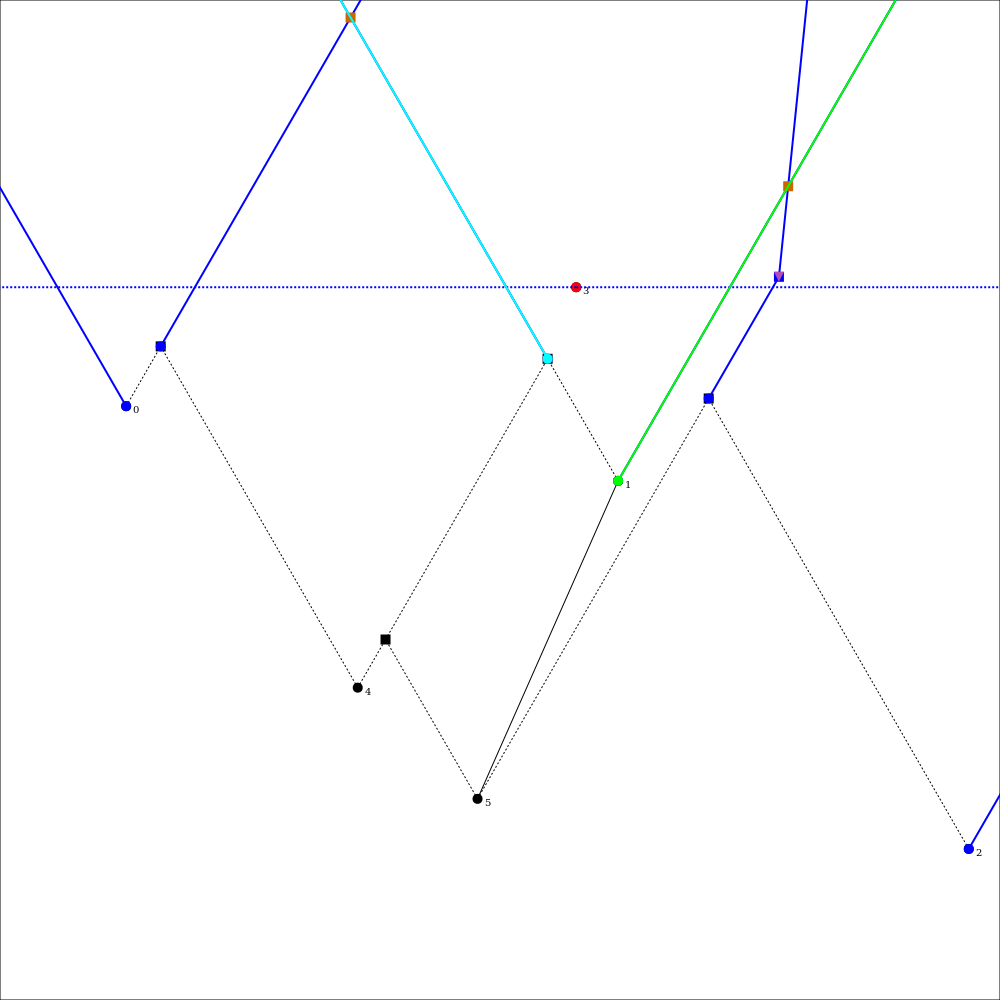}
	\end{subfigure}
	\begin{subfigure}{\execfigwidth}
	\centering
	\includegraphics[width=\textwidth]{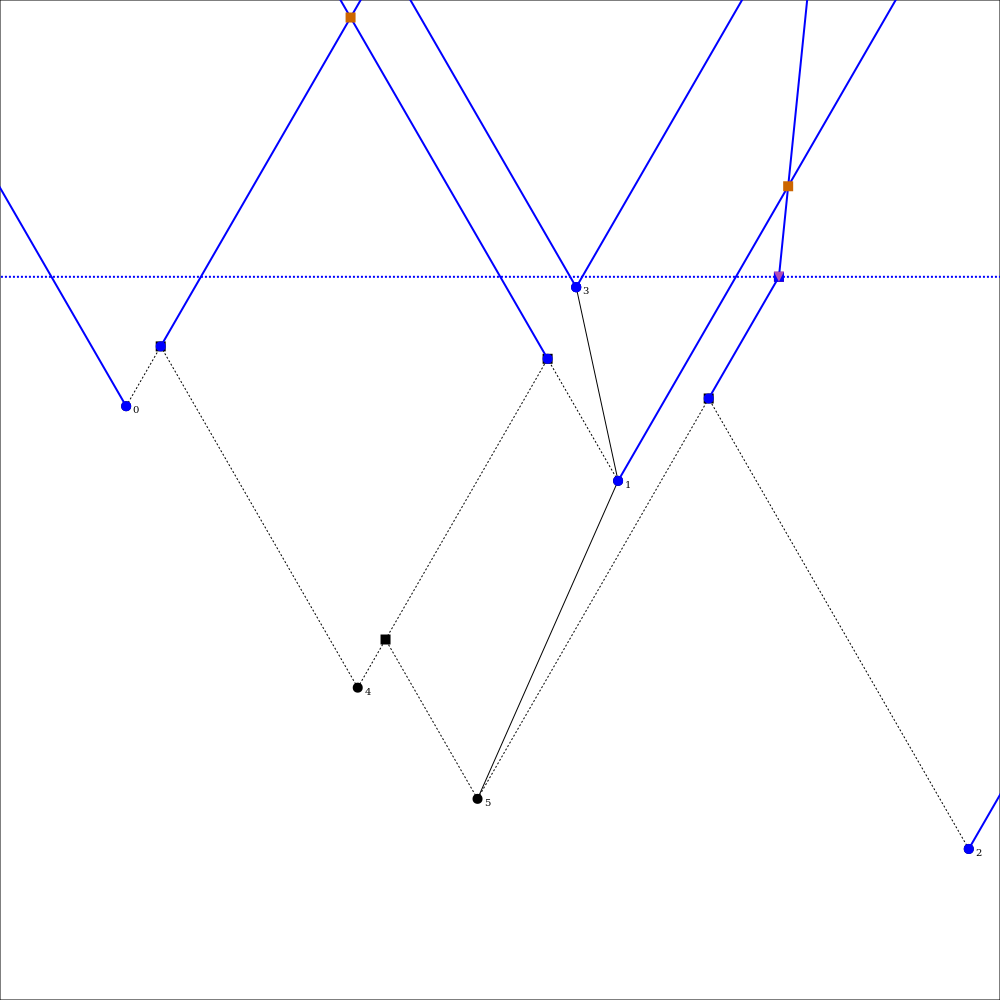}
	\end{subfigure}
	\begin{subfigure}{\execfigwidth}
	\centering
	\includegraphics[width=\textwidth]{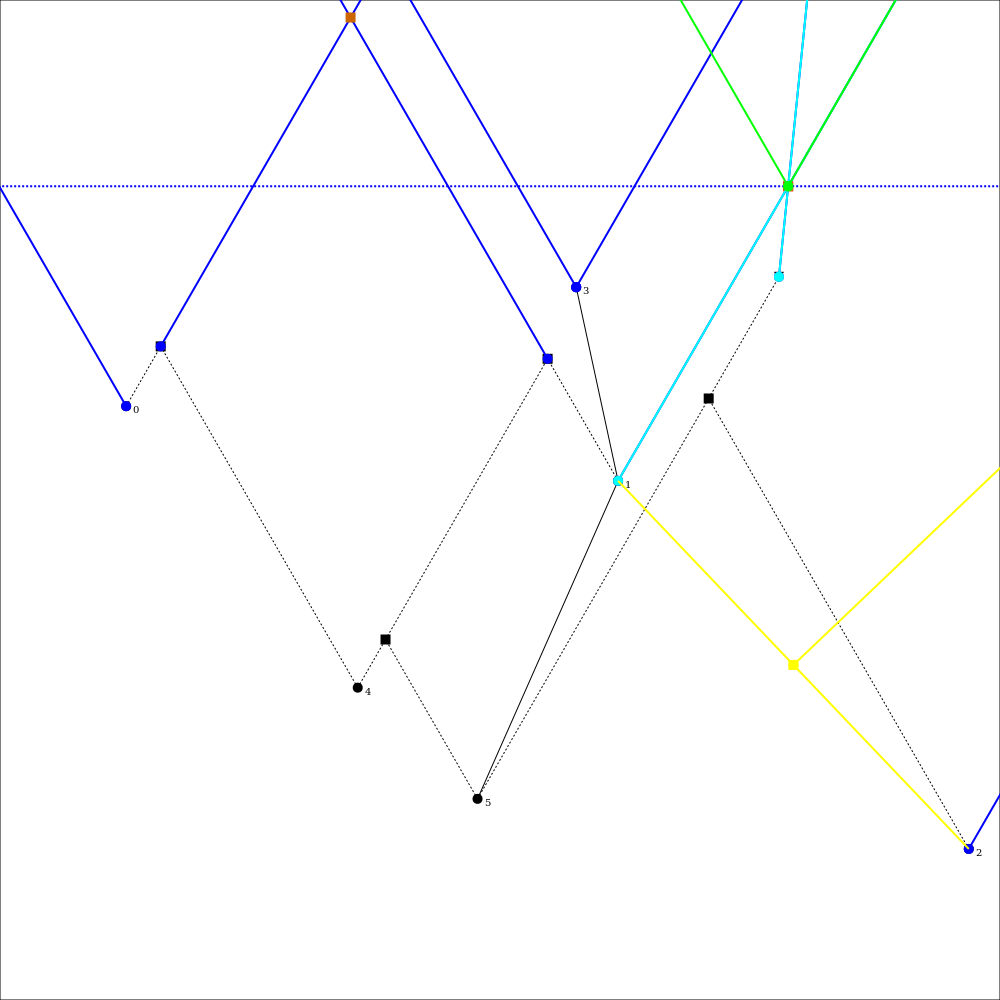}
	\end{subfigure}
	\begin{subfigure}{\execfigwidth}
	\centering
	\includegraphics[width=\textwidth]{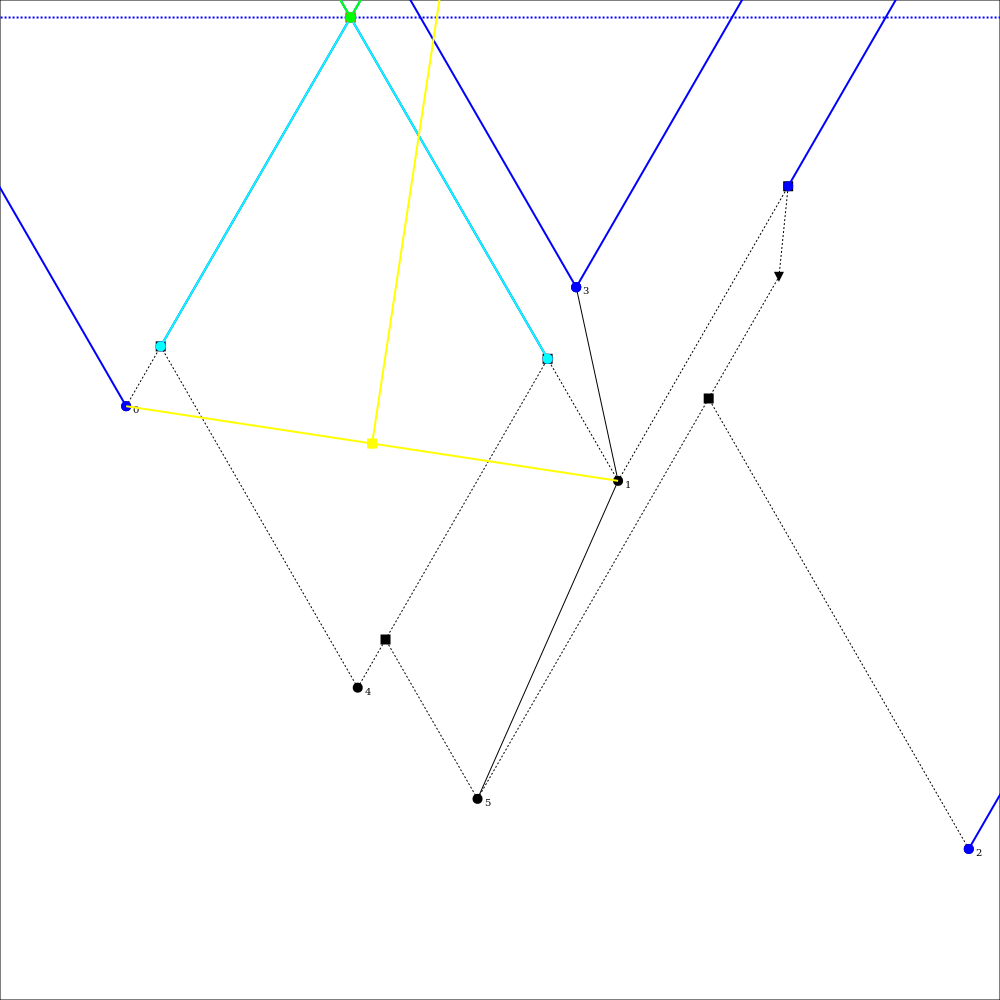}
	\end{subfigure}
	\begin{subfigure}{\execfigwidth}
	\centering
	\includegraphics[width=\textwidth]{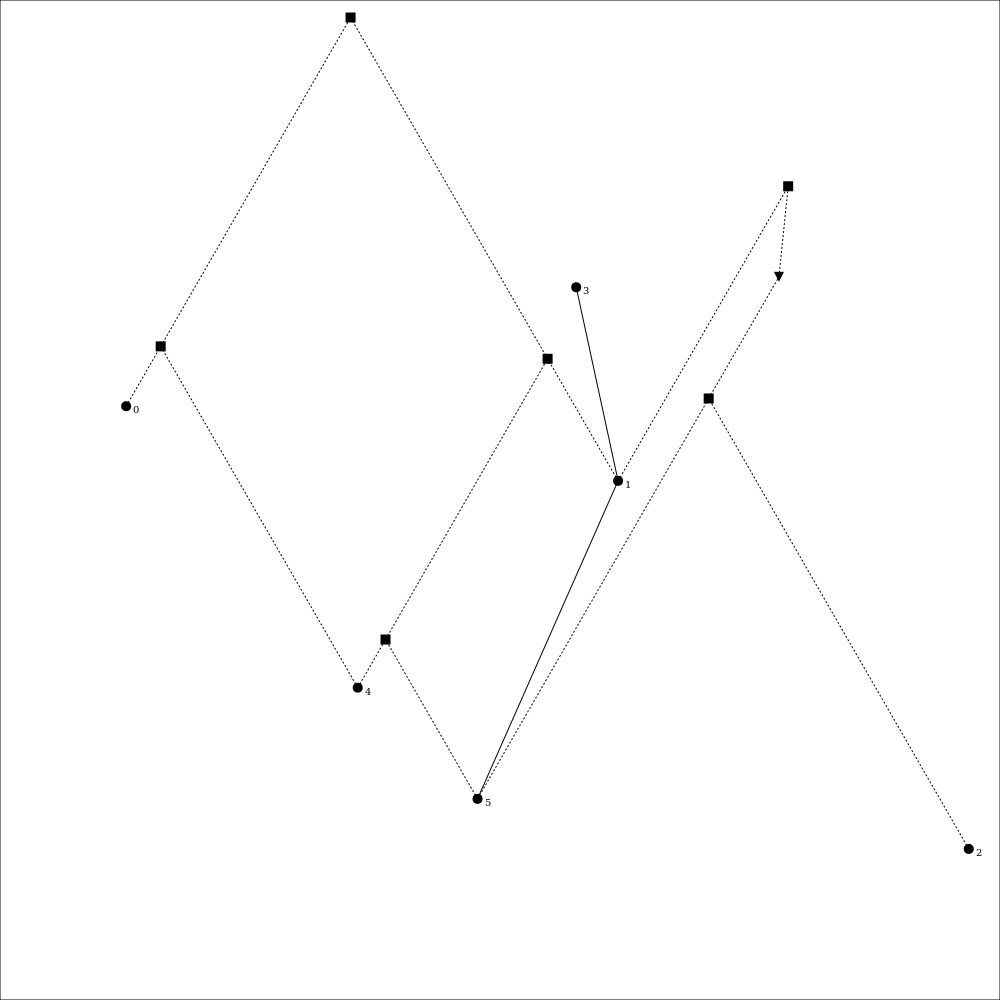}
	\end{subfigure}
	\caption{Sample execution for $n = 6$ points and cone $\mathcal{C} = (\nicefrac{4\pi}{3}, \nicefrac{5\pi}{3})$,
		resulting in $\tau = \nicefrac{\pi}{2}$ (upward).
		The currently processed point is marked in red, the sweepline is a dashed, blue line.
		All rays currently intersecting the sweepline are blue, except for the left boundary ray $B_L$ (cyan) and right boundary ray $B_R$ (green).
		Intersection points are marked as squares, deletion points as triangles.
		For intersection events, the intersecting rays are cyan and the the bisector line is yellow.
		Edges of the Yao graph are solid black lines and settled cone boundaries are dashed.}
	\label{fig:app:example}
\end{figure}

\FloatBarrier
\newpage
\subsection{Results}

\newlength{\runtimefigwidth}
\setlength{\runtimefigwidth}{\textwidth}

\begin{figure}[tbh]
	\centering
	\captionsetup[subfigure]{justification=centering}
	\begin{subfigure}[t]{\runtimefigwidth}
		\centering
		\includegraphics[width=\textwidth]{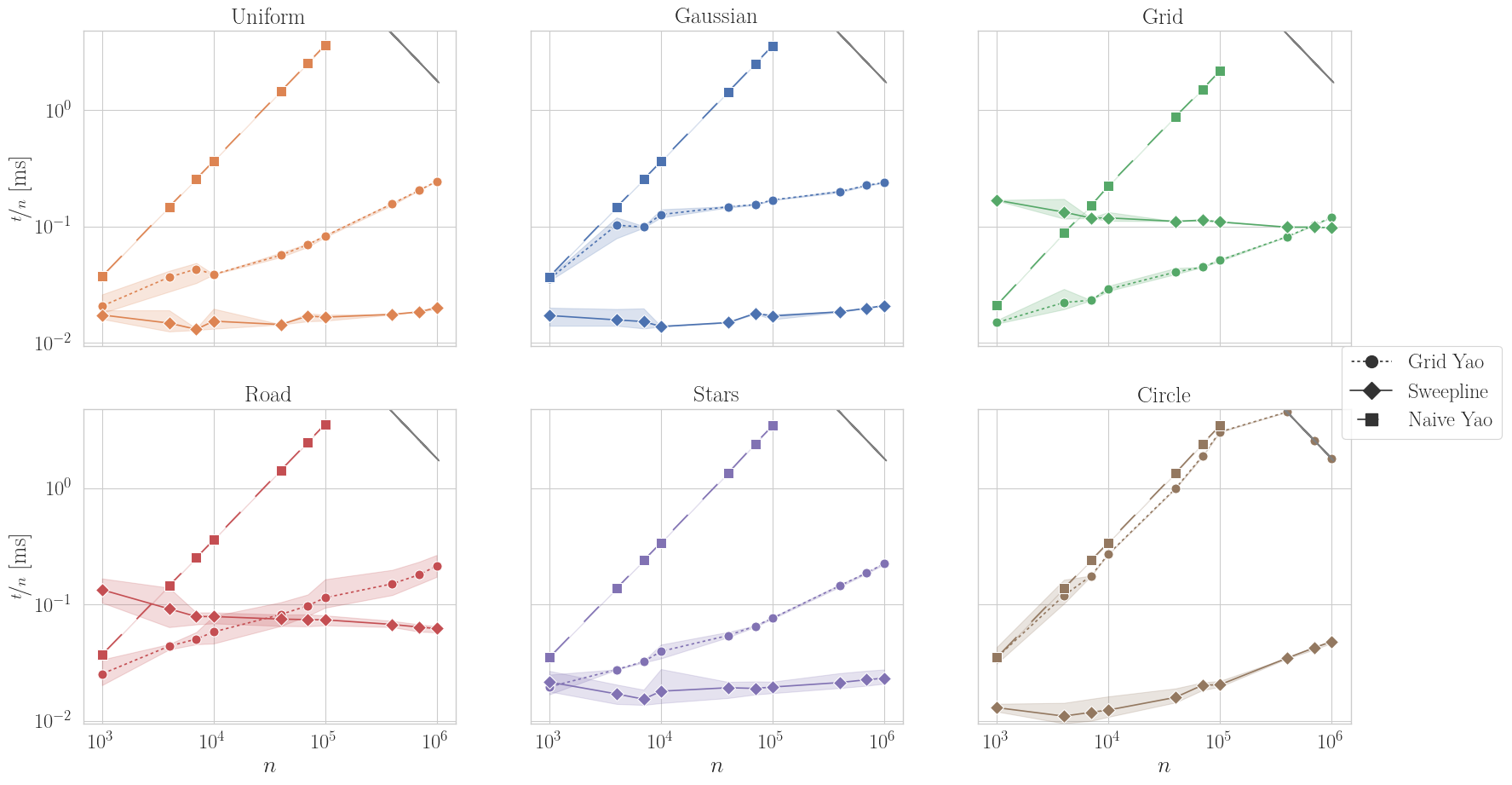}
		\caption{Inexact kernel.}
		\label{fig:app:runtimeInexact}
	\end{subfigure}

	\begin{subfigure}[t]{\runtimefigwidth}
		\centering
		\includegraphics[width=\textwidth]{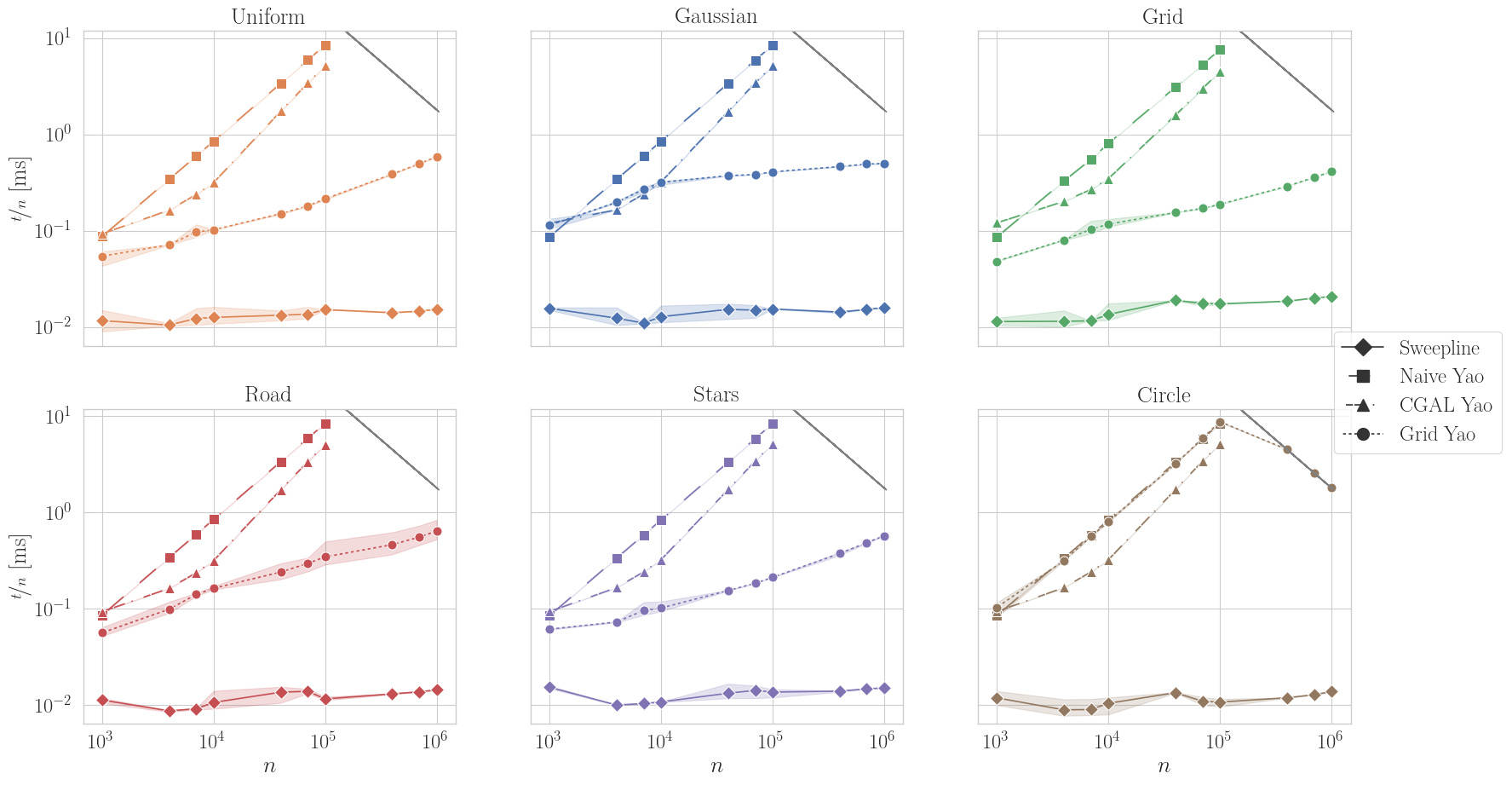}
		\caption{CGAL EPIC kernel.}
		\label{fig:app:runtimeCGALInexact}
	\end{subfigure}
\end{figure}

\begin{figure}[tbh]\ContinuedFloat
	\centering
	\captionsetup[subfigure]{justification=centering}
	\begin{subfigure}[t]{\runtimefigwidth}
		\centering
		\includegraphics[width=\textwidth]{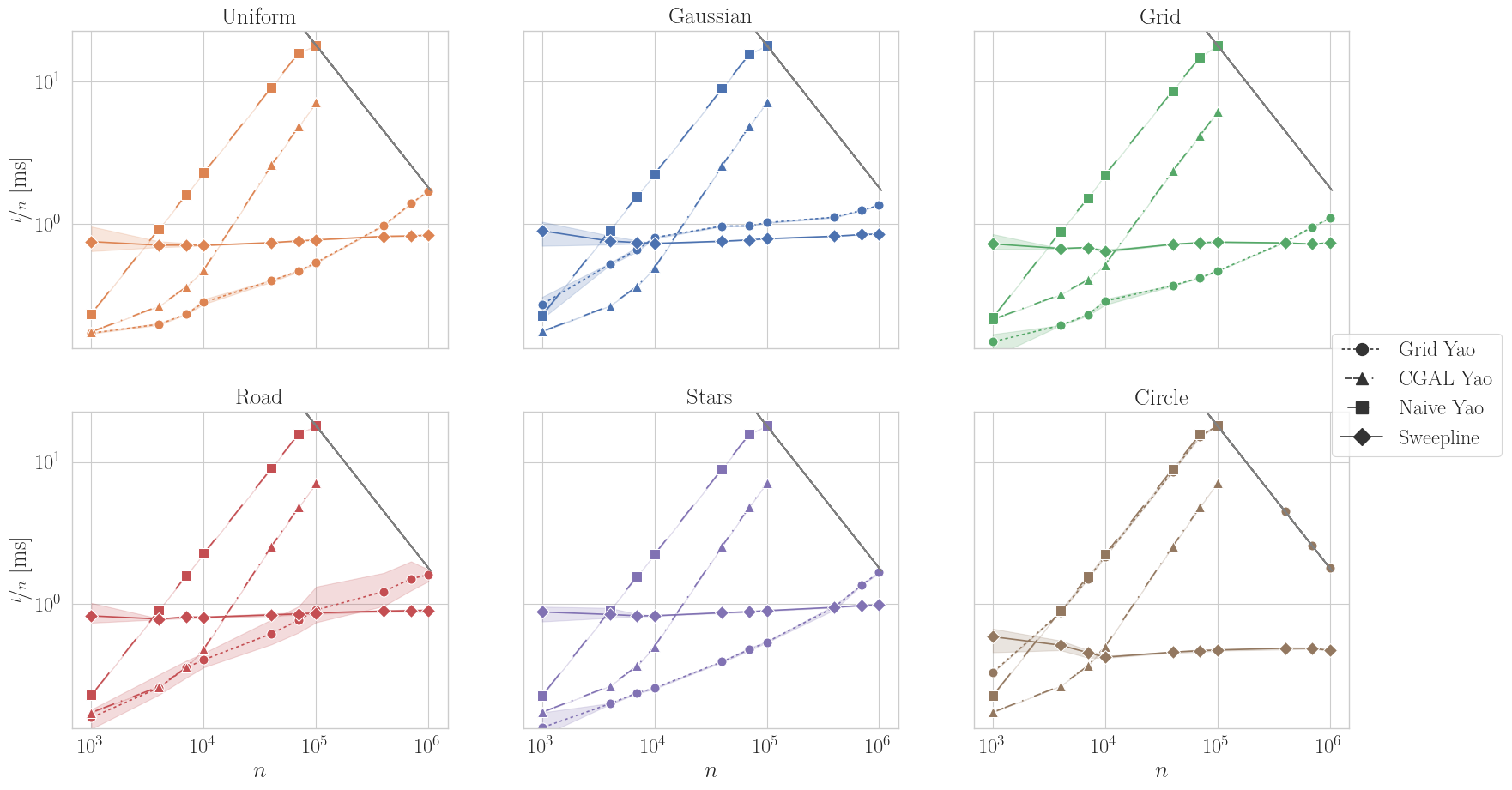}
		\caption{CGAL EPEC kernel.}
		\label{fig:app:runtimeCGALExact}		
	\end{subfigure}
	\caption{Algorithm runtime experiments.
		Experiments over varying input sizes are performed with $k = 6$ cones.
		The gray line represents the time limit of \qty{30}{\minute} per algorithm.}
	\label{fig:app:algRuntime}
\end{figure}

\end{document}